\documentclass[aps,prd,showpacs,twocolumn,
superscriptaddress]{revtex4-1}

\usepackage{graphicx}
\usepackage{dcolumn}

\usepackage{amsmath}
\usepackage{amssymb}
\usepackage{bm}
\usepackage{color}
\usepackage[normalem]{ulem}
\usepackage[dvipsnames]{xcolor}
\usepackage{hyperref}
\hypersetup{
colorlinks=true, 
linkcolor=blue,  
citecolor=cyan,  
}

\begin{document}
\title{Electromagnetic field around boosted rotating black hole }

%
\author{Ahmadjon Abdujabbarov}
\email{ahmadjon@astrin.uz}
\affiliation{Center for Field Theory and Particle Physics and Department of Physics, Fudan University, 2005 Songhu Road, 200438 Shanghai, China}
\affiliation{Ulugh Beg Astronomical Institute, Astronomicheskaya 33,
Tashkent 100052, Uzbekistan }

\author{Naresh~Dadhich}\email{nkd@iucaa.in}

\affiliation{Inter-University Centre for Astronomy and Astrophysics, Post Bag 4, Ganeshkhind,
Pune 411 007, India }

\author{Bobomurat Ahmedov}\email{ahmedov@astrin.uz}

\affiliation{Ulugh Beg Astronomical Institute, Astronomicheskaya 33,
Tashkent 100052, Uzbekistan }
\affiliation{National University of Uzbekistan, Tashkent 100174, Uzbekistan}

\begin{abstract}
The exact analytic solutions of the Maxwell equations in the exterior of a boosted rotating black hole immersed in an external magnetic field are obtained. The effect of boost as well as electromagnetic field on charged particle motion and energy extraction process -- magnetic Penrose process around black hole is studied. It justifies however the well known statement that magnetic field greatly increases efficiency of energy extraction which is further boosted by the boost velocity.

\end{abstract}

\maketitle

\section{Introduction}

According to no-hair theorem a black hole cannot have its own magnetic field.  It could though have an electric charge which is however not sustainable astrophysically because it would attract opposite charge from surrounding region and would get neutralized. An accreting medium could however produce magnetic field due to currents in accretion disk. Due to theremarkable frame dragging phenomena of rotating black hole, magnetic field lines would get twisted and that in turn would produce quadrupole electric field. It is that which is really responsible for the well known and highly efficient energy extraction processes  -- Blandford-Znajek and magnetic Penrose process (MPP) \cite{Wald74,Blandford1977,Dhurandhar83,Dhurandhar84,Dhurandhar84b,wagh85,Dadhich18}.

The first study of electromagnetic field around rotating black hole was done long back by Wald in 1974~\cite{Wald74}.  He considered a rotating black hole sitting in an  asymptotically uniform  magnetic field. As mentioned earlier, the frame dragging effect due to rotation of the black hole creates additional quadrupole electric field around the black hole. The structure and properties of the electromagnetic field surrounding the black hole is studied in various papers~\cite{Petterson74,Chitre75,Dhurandhar83,Dhurandhar84,Dhurandhar84b,wagh85,Dadhich18,Aliev02,Aliev2004,Abdujabbarov08,Karas09,Abdujabbarov10,Frolov10,Abdujabbarov11a,Frolov12,Karas12,Zahrani13,Abdujabbarov13a,Abdujabbarov13b,
Morozova14,Abdujabbarov14,Ahmedov12,Hakimov13,Stuchlik14a,Nathanail17,Karas12a,Stuchlik16,Rayimbaev16}.

 Like all other astrophysical objects, black holes also move as we have the phenomena of black hole mergers producing gravitational waves which have been for the first time observed in the most remarkable discovery of our times ~\cite{LIGO16b,LIGO16c,LIGO16d}. This wonderfully underlines the important role boost velocity can play in the merger process as well as other physical processes around black holes.
The linear momentum of the black hole can be tested using the analysis of possible observation of the electromagnetic counterpart from black hole merger ~\cite{Morozova14,Lyutikov11}.
This fact indicates the importance of the inclusion of the boost parameter to Kerr spacetimes in order to study the effects of the boost velocity to the geometry (gravitational field) of black hole spacetime. Recently an analytic solution of Einstein equation for boosted rotating black hole has been obtained in~\cite{Soares17}. The effect of gravitational lensing around boosted black hole is considered in~\cite{Benavides18}. Here we consider the electromagnetic field  produced around it by asymptotically uniform external magnetic field.

In this work our main purpose is to study the electromagnetic field around boosted rotating black hole immersed in external uniform magnetic field. The paper is organized as follows: The  Sect.~\ref{sect2} is devoted to study the electromagnetic field while Sect.~\ref{sect3} is devoted to study of charged particle motion. The energy extraction by magnetic Penrose process has been discussed in Sect.~\ref{sect4}.
In Sect.~\ref{sect6}, we conclude the main results of the paper. Throughout the paper we use $G=c=1$ and $(-+++)$ signature. Greek indices run from 0 to 3 and Latin from 1 to 3.

\section{Boosted black hole in external magnetic field \label{sect2}}

Consider the spacetime metric around a boosted rotating black hole in Kerr-Schild coordinates $(t, r, \theta, \phi)$ given by~\cite{Soares17} 
\begin{eqnarray}
ds^2&=&  -\left( 1-\frac{2Mr}{\Sigma}\right) dt^2 + \left( 1+\frac{2Mr}{\Sigma}\right) dr^2 +\frac{\Sigma}{\Lambda} d\theta^2
\nonumber \\ &&   +\frac{A \sin^2\theta }{\Lambda^2 \Sigma} d\phi^2 - \frac{4Mr}{\Sigma} dtdr-\frac{4Mra \sin^2\theta}{\Lambda \Sigma}dtd\phi \nonumber\\&&  -\frac{2a\sin^2\theta }{\Lambda} \left(1+\frac{2Mr}{\Sigma}\right) drd\phi,
\label{metr}
\end{eqnarray}
where
\begin{eqnarray}
 \Sigma&=&r^2+a^2\left(\frac{\beta +\gamma\cos\theta}{\gamma +\beta\cos\theta}\right)^2\ ,\\
 \Lambda&=& (\gamma +\beta\cos\theta)^2\ ,\\
 A&=& \Sigma^2 \Lambda+a^2 (\Sigma+2 Mr) \sin^2\theta\ ,
\end{eqnarray}
with $a=J/M$ being the specific angular momentum of the black hole of total mass $M$, and
\begin{eqnarray}
\gamma= (1-v^2)^{-1/2}, \, \, \, \beta=v\gamma.
\end{eqnarray}

{Note, that in~\cite{Soares17} the metric of spacetime around boosted Kerr black hole is expressed in retarded Robinson-Trautman coordinates. Using the coordinate transformation for retarded time as $u=t-r$, one can obtain the metric in Kerr-Schild coordinates. }

The event horizon of the boosted black hole is the same as for Kerr black hole which is defined as the surface $g^{rr}=0$:
\begin{eqnarray}
r_{h}=M+\sqrt{M^2-a^2}\ .
\end{eqnarray}

This is because horizons (event and Cauchy) are null surfaces and hence are invariant under boost. On the other hand static limit surface $r_{\rm st}$ is not so and hence boundary of ergoregion is altered by the boost velocity.  {The static limit has the following form
\begin{eqnarray}
r_{\rm st}= M+ \sqrt{M^2 -a^2\left(\frac{\beta+\gamma \cos\theta}{\gamma+\beta \cos\theta}\right)^2}\ ,
\end{eqnarray}
which takes the following simple form in the equatorial plane
}
\begin{eqnarray}
r_{\rm st}= M+ \sqrt{M^2 -v^2a^2}\ .
\end{eqnarray}
It however coincides with the horizon on the axis, $\theta=0$. It is interesting to note that the width of the ergoregion shrinks with increasing velocity and it approaches zero as $v\to1$.  This means that large boost velocity decreases the region available for existence of negative energy orbits. It would have adverse effect on energy extraction processes, magnetic Penrose process \cite{Dhurandhar83,Dhurandhar84,Dhurandhar84b} as well as Blandford-Znajek mechanism \cite{Blandford1977} which critically depend upon negative energy orbits.

\begin{figure*}[t!]
\begin{center}
\includegraphics[width=0.24\linewidth]{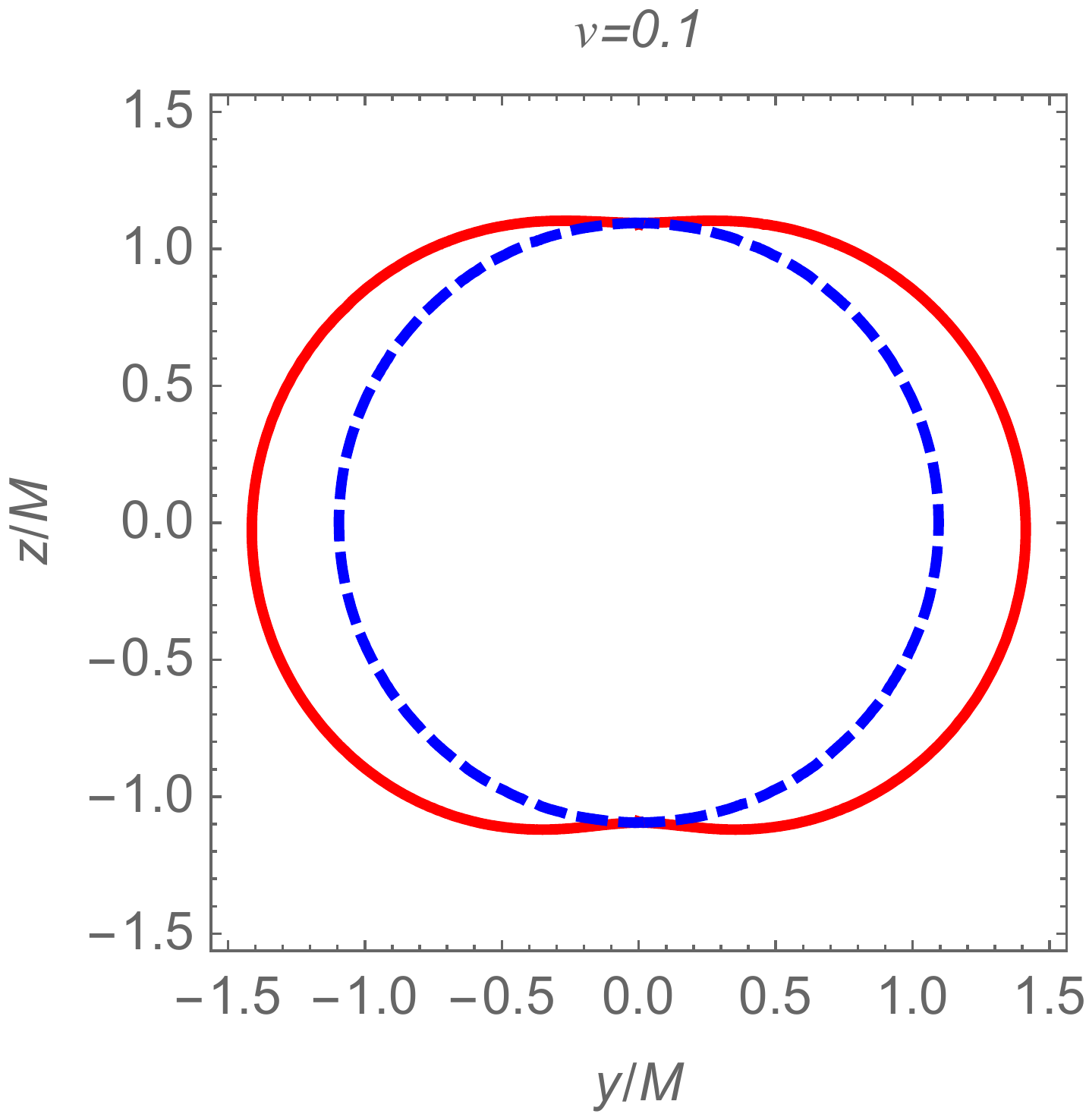}
\includegraphics[width=0.24\linewidth]{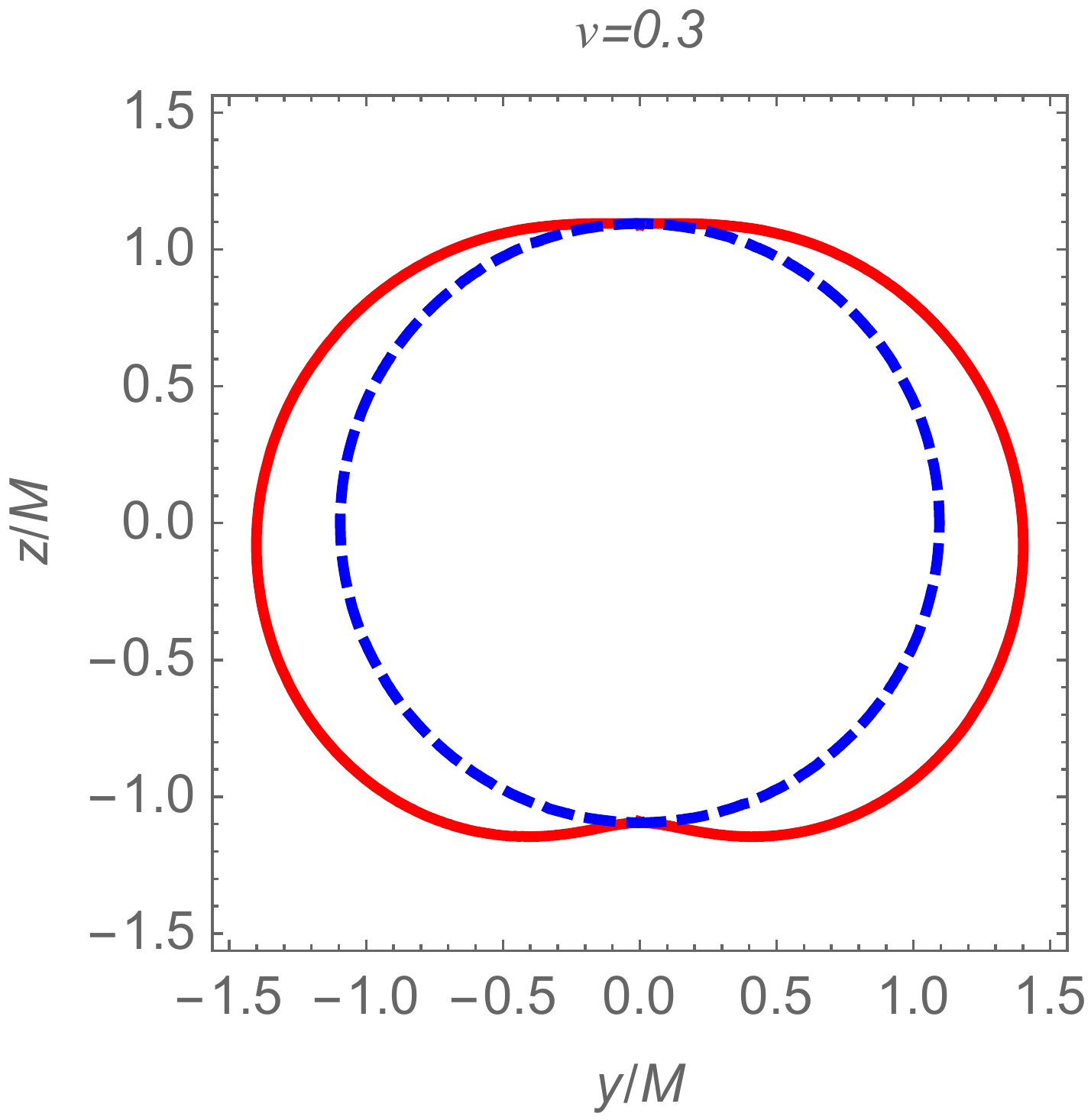}
\includegraphics[width=0.24\linewidth]{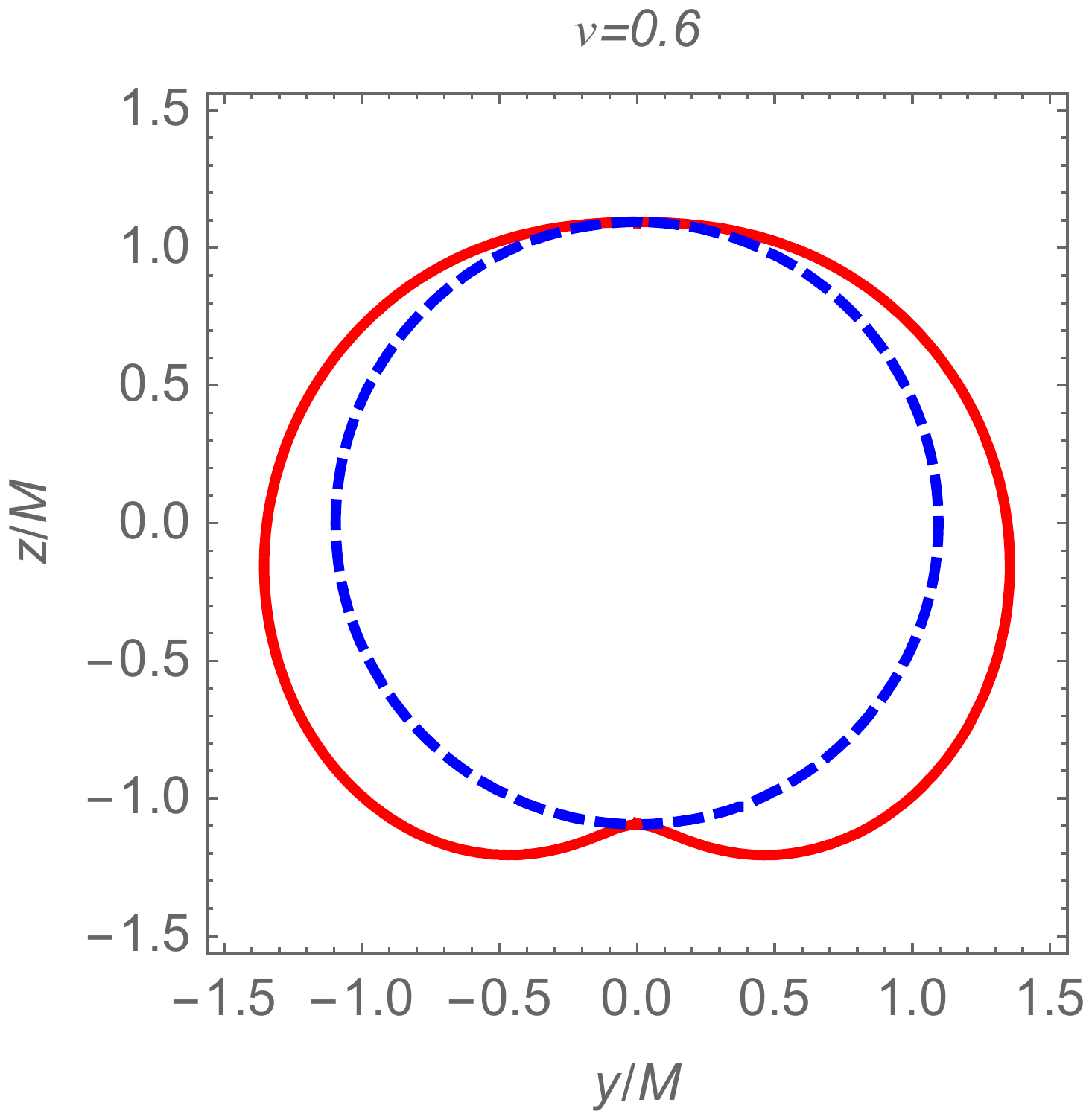}
\includegraphics[width=0.24\linewidth]{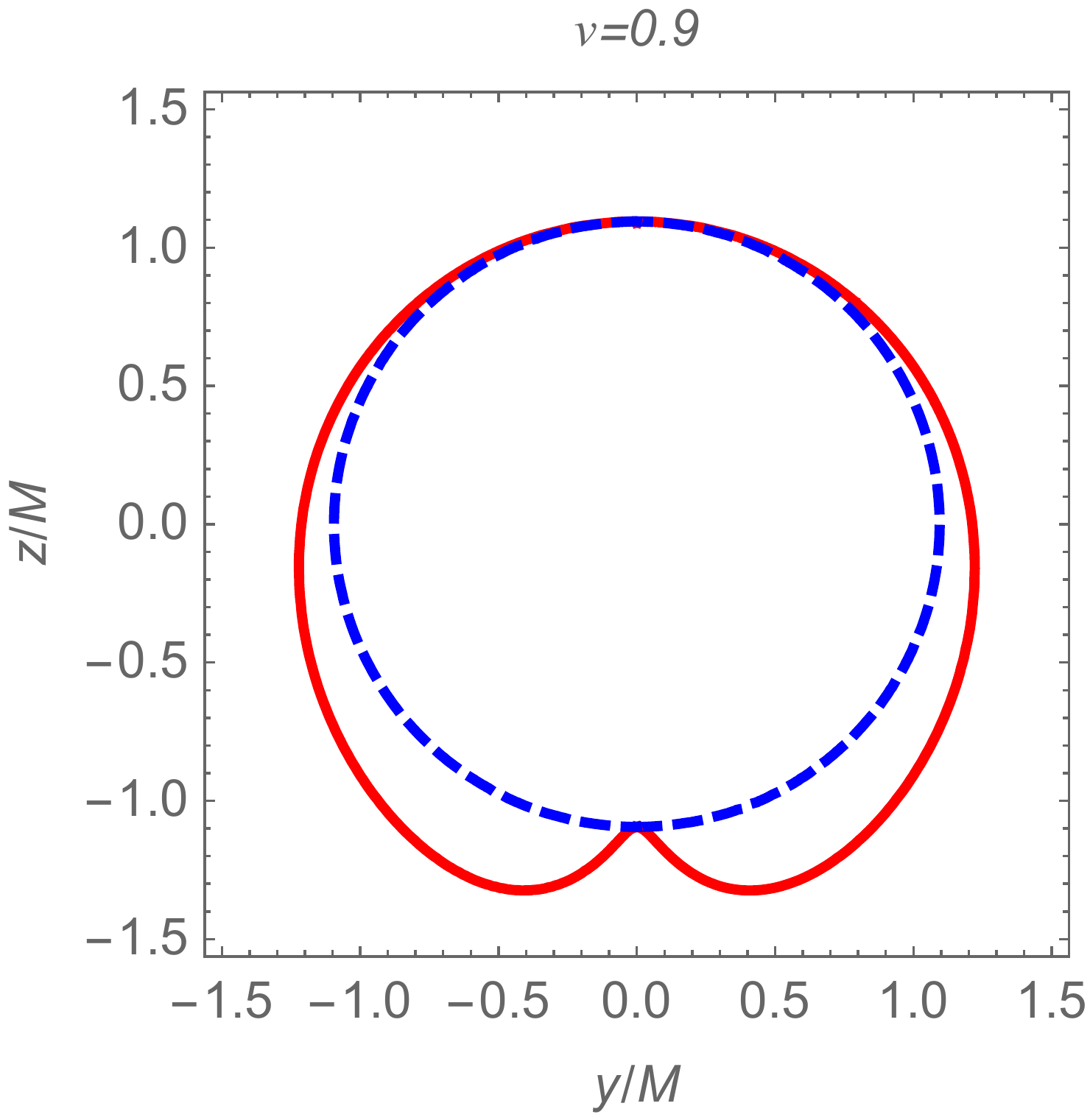}

\end{center}
\caption{The cross-section of the ergoregion in the $y-z$ plane for the different values of the boost velocity. The rotation parameter has been taken to  be $a/M=0.9$. The dashed line is the event horizon and solid line is the static limit.   \label{ergoreg}}
\end{figure*}

 In the Fig.~\ref{ergoreg} the ergoregion is shown to depict the effect of boost velocity on its shape. The boost is along the positive $z$ direction and that is why the static limit gets smeared at the front end while it develops a dip at the back end which deepens with increase in velocity.

Now we consider the electromagnetic field around boosted rotating black hole immersed in external asymptotically uniform magnetic field.
We will use the spacetime symmetries, especially the existence of timelike and spacelike
Killing vectors $\xi^\mu{(t)}$ and $\xi^\mu{(\phi)}$ respectively satisfying the Killing
equation
\begin{eqnarray}
\label{ke} \xi_{\mu ;\nu}+\xi_{\nu;\mu}=0\ ,
\end{eqnarray}
as well as the wave-like equation in vacuum spacetime
\begin{eqnarray}
\Box{\xi^\mu}=0\ .
\end{eqnarray}
The solution of vacuum Maxwell equations $\Box A^\mu = 0$ for the vector potential $A_\mu$ of the electromagnetic field in the Lorentz gauge can be expressed as linear combnation of the Killing vectors in the following form~\cite{Wald74}
\begin{equation}
A^\mu=C_1 \xi^\mu{(t)}+C_2 \xi^\mu{(\phi)}\ ,
\end{equation}
where the constants $C_1$ and $C_2$ are to be found by the using the asymptotic behavior of spacetime and electromagnetic field.  Asymptotically magnetic field attains its uniform value $B$ and thereby fixing $C_2=B/2$. In order to find the value of the constant $C_1$ one can use the the asymptotic properties of spacetime
(\ref{metr}) at the infinity as well as the  electrical neutrality of the black hole
\begin{eqnarray}
\label{flux} 4\pi Q=0&=& \frac{1}{2}\oint
F^{\mu\nu}{_*dS}_{\mu\nu}\nonumber\\ &=&C_1 \oint
\Gamma^\mu_{\nu\rho}u_\mu m^\nu \xi^\rho{(t)}(uk)dS
\nonumber \\ && +\frac{B}{2}\oint \Gamma^\mu_{\nu\rho}u_\mu m^\nu
\xi^\rho_{(\phi)}(uk)dS\ ,
\end{eqnarray}
where integral is taken through the surface
at the asymptotic infinity. Using the asymptotic values of the Christoffel symbols  one can easily get the value of constant
$$C_1= aBk\ ,$$
where 
\begin{eqnarray}
k= aB \frac{3}{\beta^2} +\frac{3\gamma}{2 \beta^3} \log\frac{\gamma-\beta}{\gamma+\beta} .
\end{eqnarray}
For $v=0$, $k=1$ while for $v=1$, $k=0$.

Finally the 4-vector potential $A_\mu$ of the electromagnetic
field will take the form
\begin{eqnarray}
A_0&=& a B \frac{ k \Lambda ( 2 M r - \Sigma) -
    M r \sin^2\theta}{\Lambda \Sigma }\\
A_1&=& a B \frac{ 4 k M r \Lambda - (2 M r + \Sigma) \sin^2\theta}{2 \Lambda\Sigma} \\
A_3&=& B  \frac{ \Sigma^2 -4 a^2 k M r \Lambda +
    a^2 (2 M r + \Sigma) \sin^2\theta}{2 \Lambda^2 \Sigma \sin^{-2}\theta}
\end{eqnarray}
Using the zero angular momentum observer (ZAMO)
\begin{eqnarray}
u^{\alpha} &\equiv & \left( \frac{1}{\sqrt{-g_{00}}},\ 0,\ 0,\ 0\right),\ \\
u_{\alpha} &\equiv & \left( \sqrt{-g_{00}},\ \frac{g_{01}}{\sqrt{-g_{00}}},\ 0,\ \frac{g_{03}}{\sqrt{-g_{00}}}\right),
\end{eqnarray}
one can find the electromagnetic field components as given by
\begin{widetext}
\begin{eqnarray}
E^{\hat r} &=&   \frac{ a B M \Lambda(2 r^2 - \Sigma) (2 k \Lambda - \sin^2\theta) }{(r^2 \gamma^2 +
    a^2 \beta^2 +
    2 (a^2 + r^2) \gamma \beta \cos\theta + (a^2 \gamma^2 +
       r^2 \beta^2) \cos^2\theta)^2} \sqrt{
   1 + \frac{a^2 \sin^2\theta}{\Sigma - 2 M r}}\ , \label{Er}\\
    E^{\hat \theta}&=&
   \frac{a B M r [r^2 -
   a^2 (2 k - 1 )]  [2 (\gamma^2 + \beta^2) \cos\theta + \gamma \
\beta (3 + \cos 2 \theta)] \sin\theta}{[r^2 \gamma^2 +
  a^2 \beta^2 +
  2 (a^2 + r^2) \gamma \beta \cos\theta + (a^2 \gamma^2 +
     r^2 \beta^2) \cos\theta^2]^2} \sqrt{\frac{\Lambda}{\Sigma - 2 M r}}\ , \label{Et}\\
 E^{\hat \phi}&=&0 \ , \\
B^{\hat r} &=& \frac{B}{\Lambda^2 \Sigma^4}\sqrt{\frac{\Sigma+2Mr}{\Sigma-2 M r}} (4 a^2 M^2 r^2 \sin^2\theta (-\Lambda^{3/2}
\Sigma \cos\theta -
      2 a^2 k \Lambda (\beta + \gamma \cos\theta) (\beta^2-\gamma +  \gamma \beta \cos\theta) \nonumber \\&& -
 \Lambda \Sigma \beta \sin^2\theta - a^2 (\beta + \gamma \cos\theta) \sin^2\theta) + (2 M
r - \Sigma) (-a^2 \Lambda \Sigma (2 M r + \Sigma) \beta \cos^2\theta \sin^2\theta \nonumber \\&&  + \beta \sin^2\theta (-2 \Lambda \Sigma^3 + a^2 \Sigma (4 k \Lambda^2 M r + \Sigma) + 4 a^3 k \Lambda^2 M r (\gamma - \beta^2) -2 a^2 (a^2 M r + \Lambda \Sigma (2 M r + \Sigma)) \sin^2\theta) \nonumber \\&& +
\cos\theta (\Lambda^{3/2}
           \Sigma (4 a^2 k \Lambda M r - \Sigma^2) +
         a^2 (-\Lambda^{3/2} \Sigma (2 M r + \Sigma) + \Sigma^2 \gamma -
\Lambda \Sigma (2 M r + \Sigma) \gamma\nonumber \\&&   + 4 a k \Lambda^2 M r \gamma (\gamma - 2 \beta^2)) \sin^2\theta - 2 a^4 M r \gamma \sin^4\theta) -
      a^3 k \Lambda^2 M r \gamma^2 \beta \sin^2 2 \theta))\ , \label{Br}\\
B^{\hat \theta} &=& \frac{B \sin\theta}{\Lambda \Sigma^3 \sqrt{\Sigma-2 M r}}\left[ r \Sigma^2 (\Sigma-2 M r) -2 a^2 k \Lambda M (4 M r - \Sigma) (2 r^2 - \Sigma)
     +
     a^2 M (4 M r - \Sigma) (2 r^2 -
        \Sigma) \sin^2\theta\right]\ , \label{Bt}\\
B^{\hat \phi}&=&\frac{aB \sin\theta}{\Sigma^4 \Lambda^2}\sqrt{\frac{\Sigma^2 + a^2 (2 M r + \Sigma) \sin^2\theta}{\Sigma - 2 M r}}[(2 a^2 M r (-4 M r + \Sigma) +
     \Lambda \Sigma (-8 M^2 r^2 + \Sigma^2)) \beta\sin^2\theta   +
  2 a^2 M r  \gamma \cos\theta\nonumber\\&& \times (\Sigma -4 M r ) \sin^2\theta +
  \Lambda ((\sqrt{\Lambda} \Sigma (-8 M^2 r^2 + \Sigma^2) +
        4 a^2 k M r (4 M r - \Sigma) \gamma) \cos\theta +
     4 a^2 k M r (4 M r - \Sigma) \beta)]\ .\label{Bp}
\end{eqnarray}
\end{widetext}

In the limiting case, when $v\rightarrow 0$,   $M/r \rightarrow 0$,  $aM/r^2 \rightarrow 0$ the expressions (\ref{Er})-(\ref{Bp}) take the following form:
\begin{eqnarray}
E^{\hat r}=0\ ,  \ \ & B^{\hat r}=& B\cos\theta \ ,\\
E^{\hat \theta}=0\ , \ \  &B^{\hat \theta}=& B\sin\theta\ ,\\
E^{\hat \phi}=0\ ,  \ \ &B^{\hat \phi}=&0\ ,
\end{eqnarray}

\begin{figure*}[t!]
\begin{center}

\includegraphics[width=0.24\linewidth]{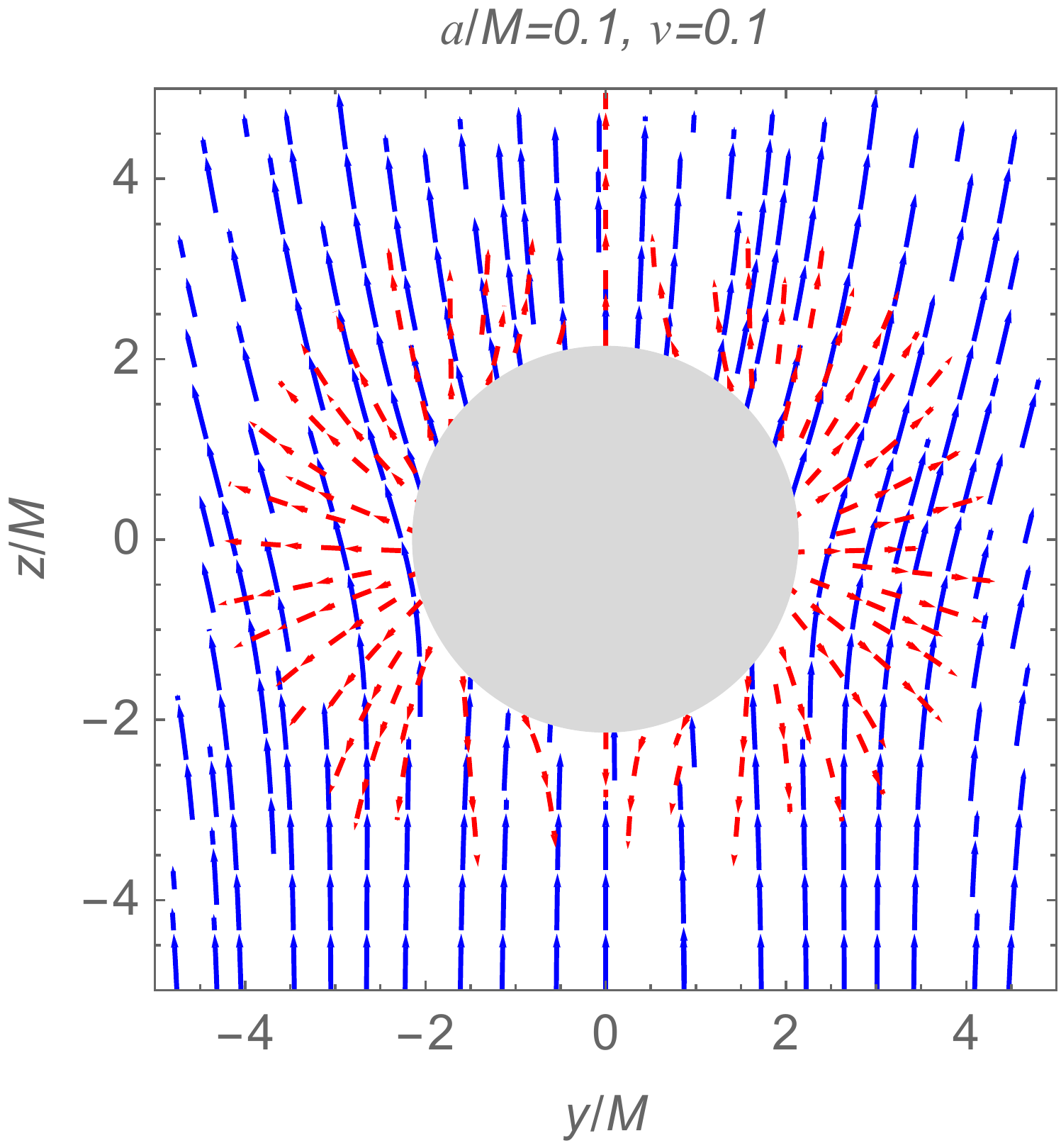}
\includegraphics[width=0.24\linewidth]{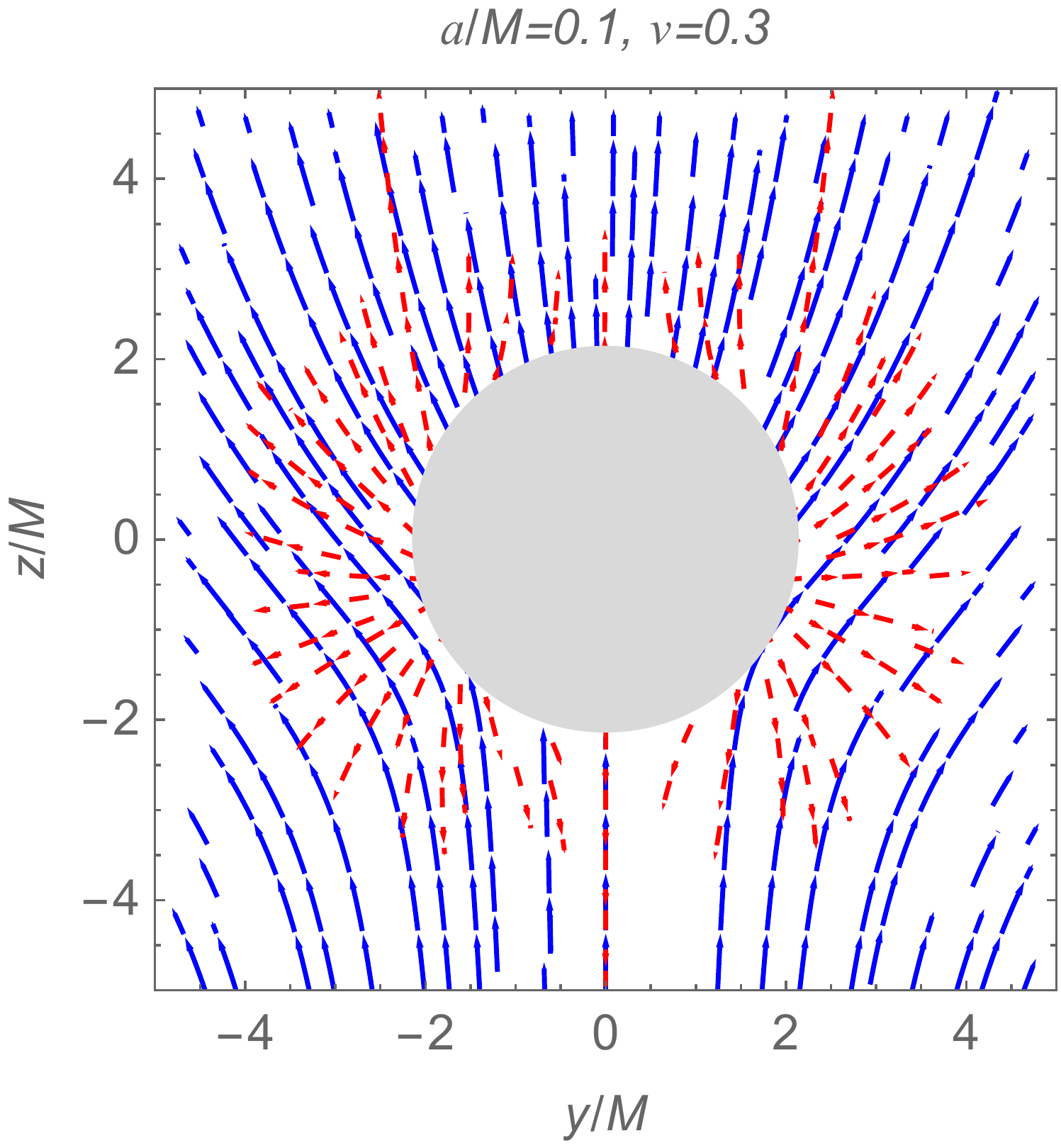}
\includegraphics[width=0.24\linewidth]{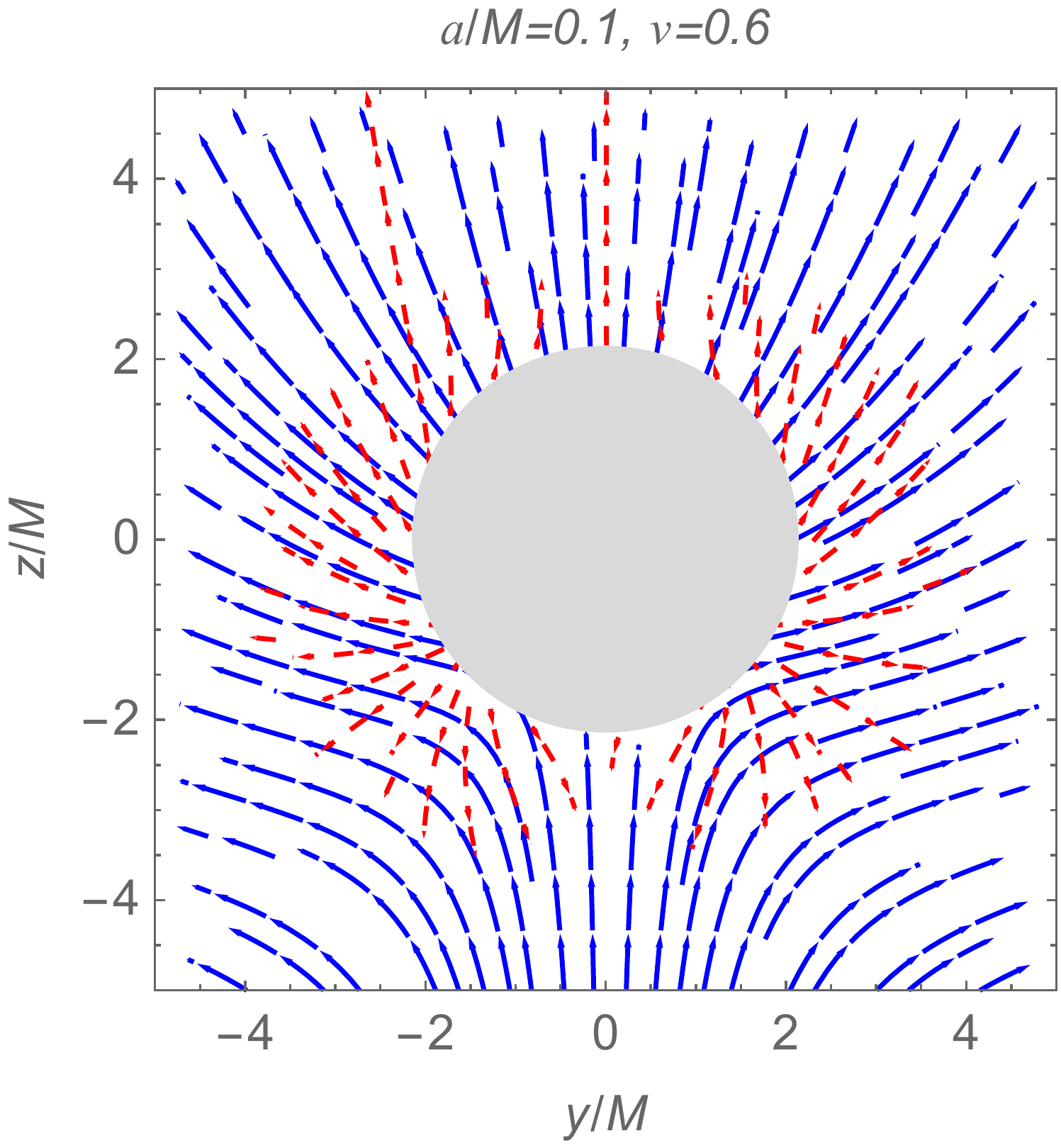}
\includegraphics[width=0.24\linewidth]{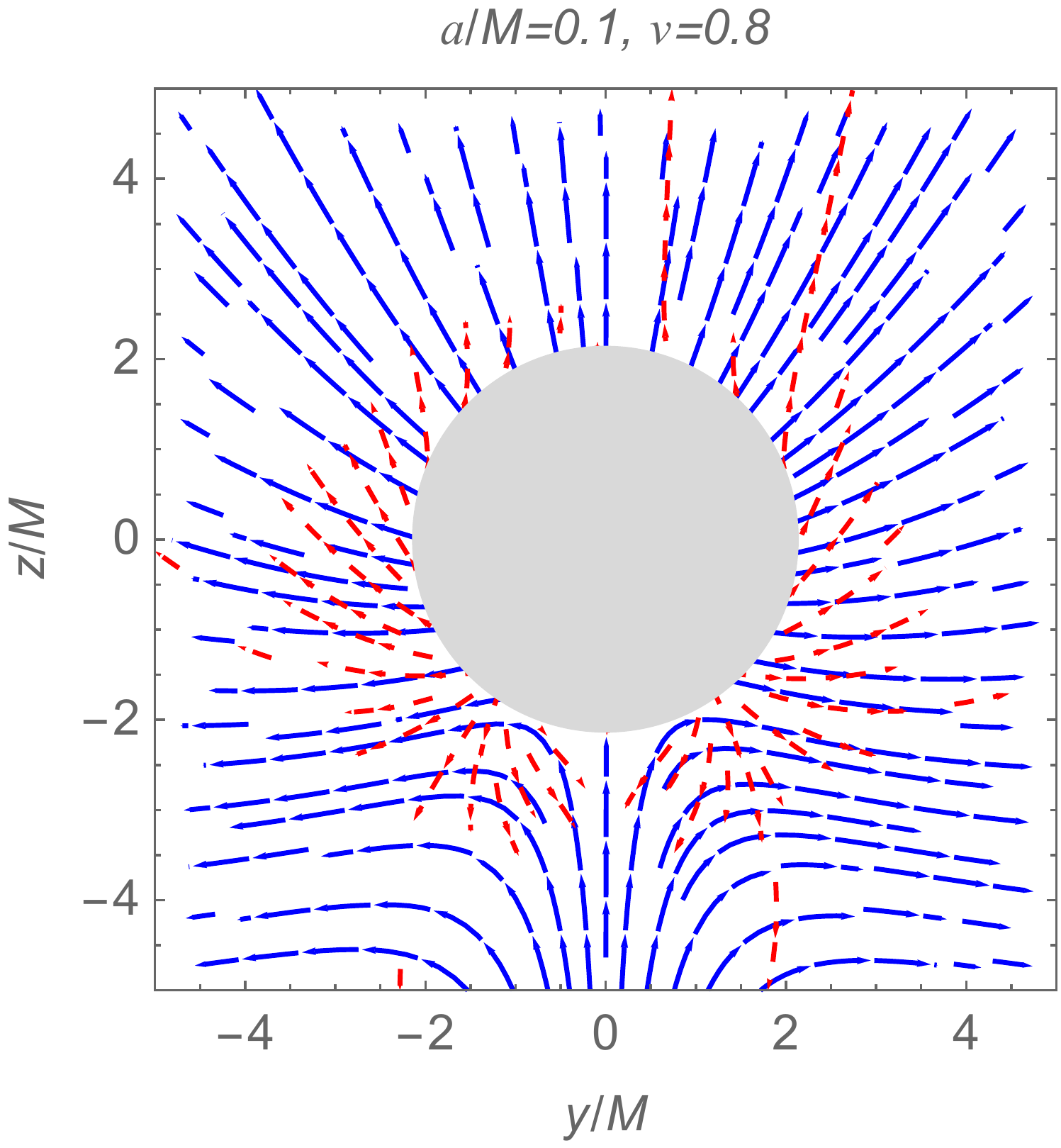}

\includegraphics[width=0.24\linewidth]{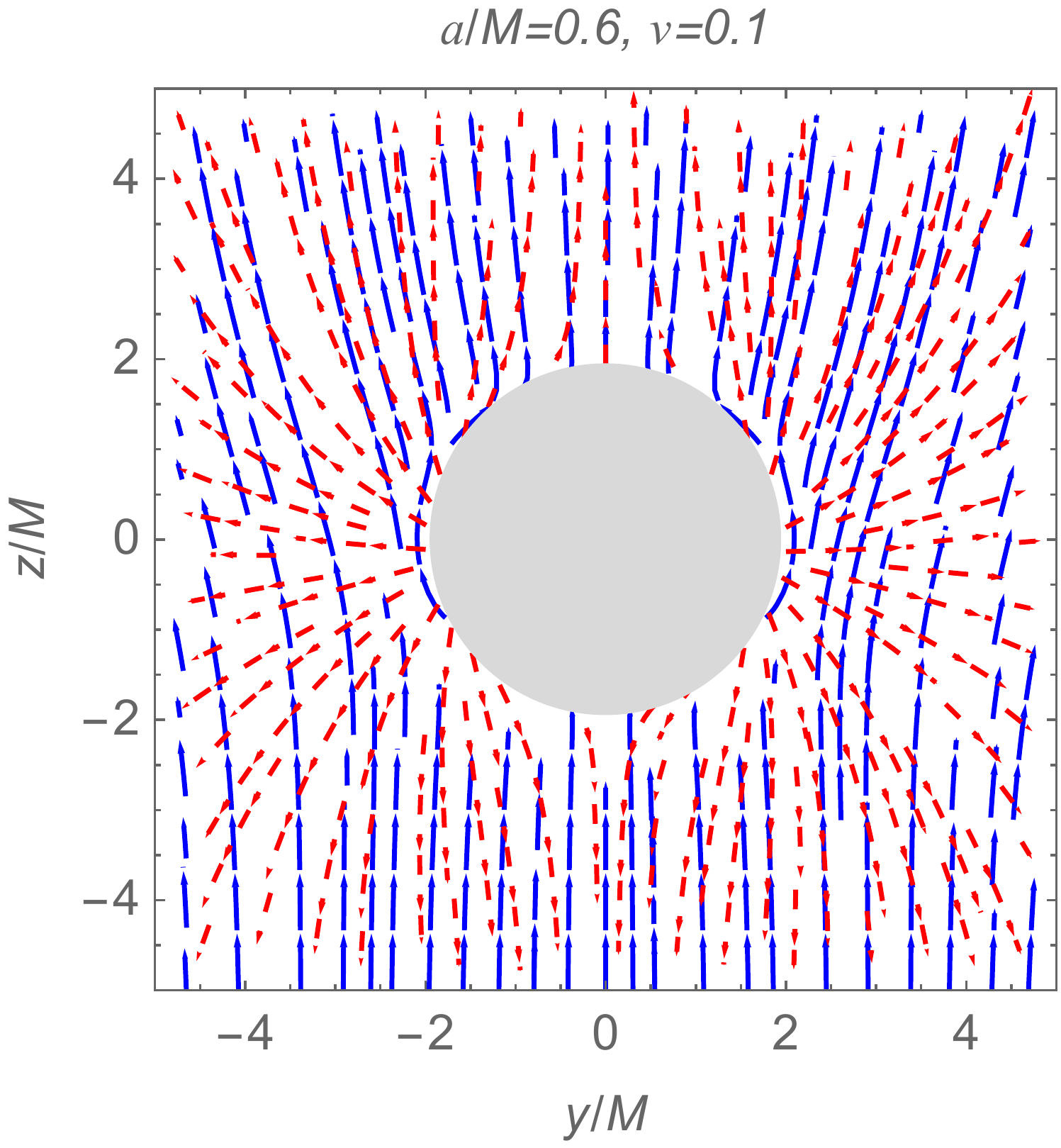}
\includegraphics[width=0.24\linewidth]{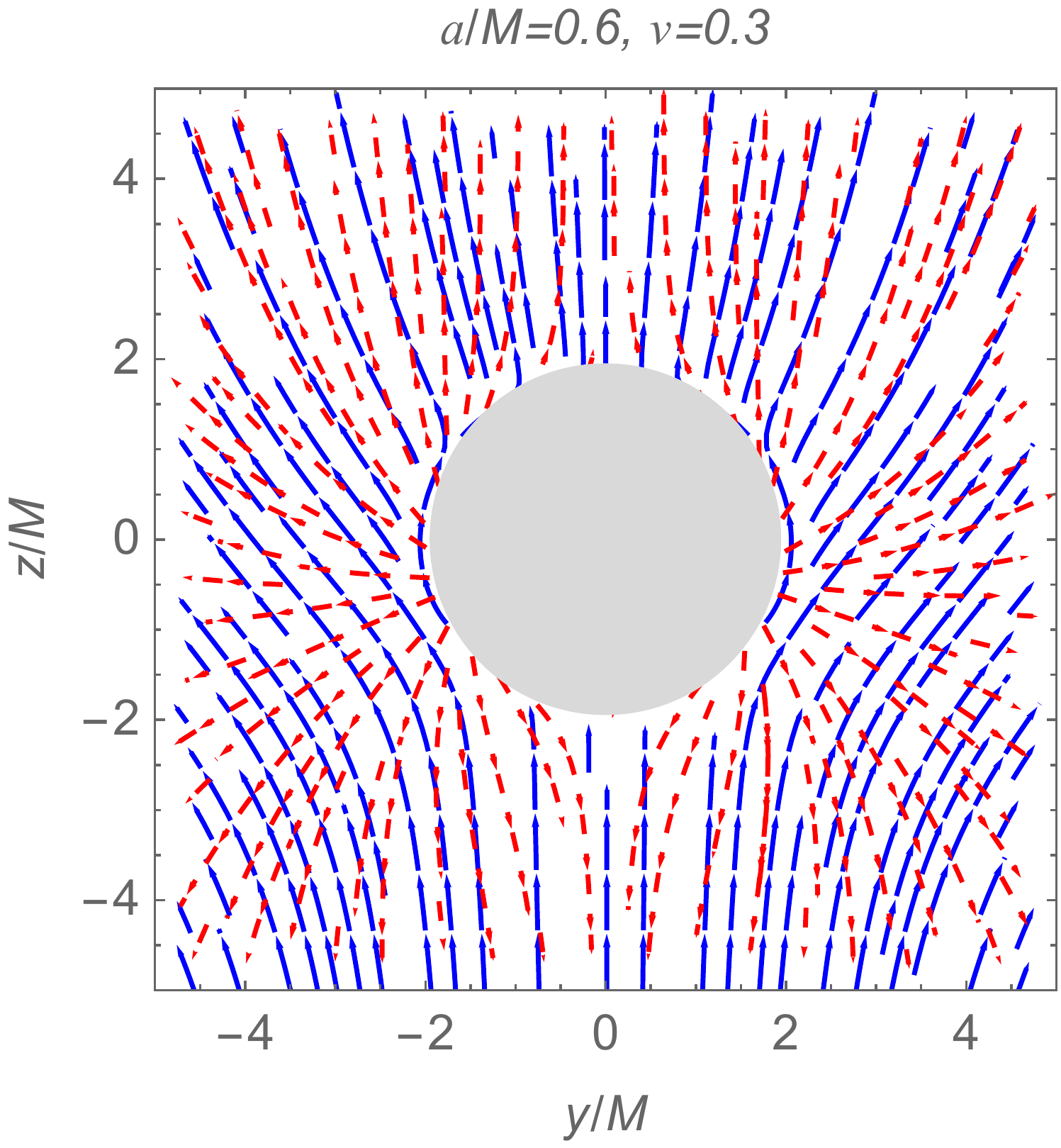}
\includegraphics[width=0.24\linewidth]{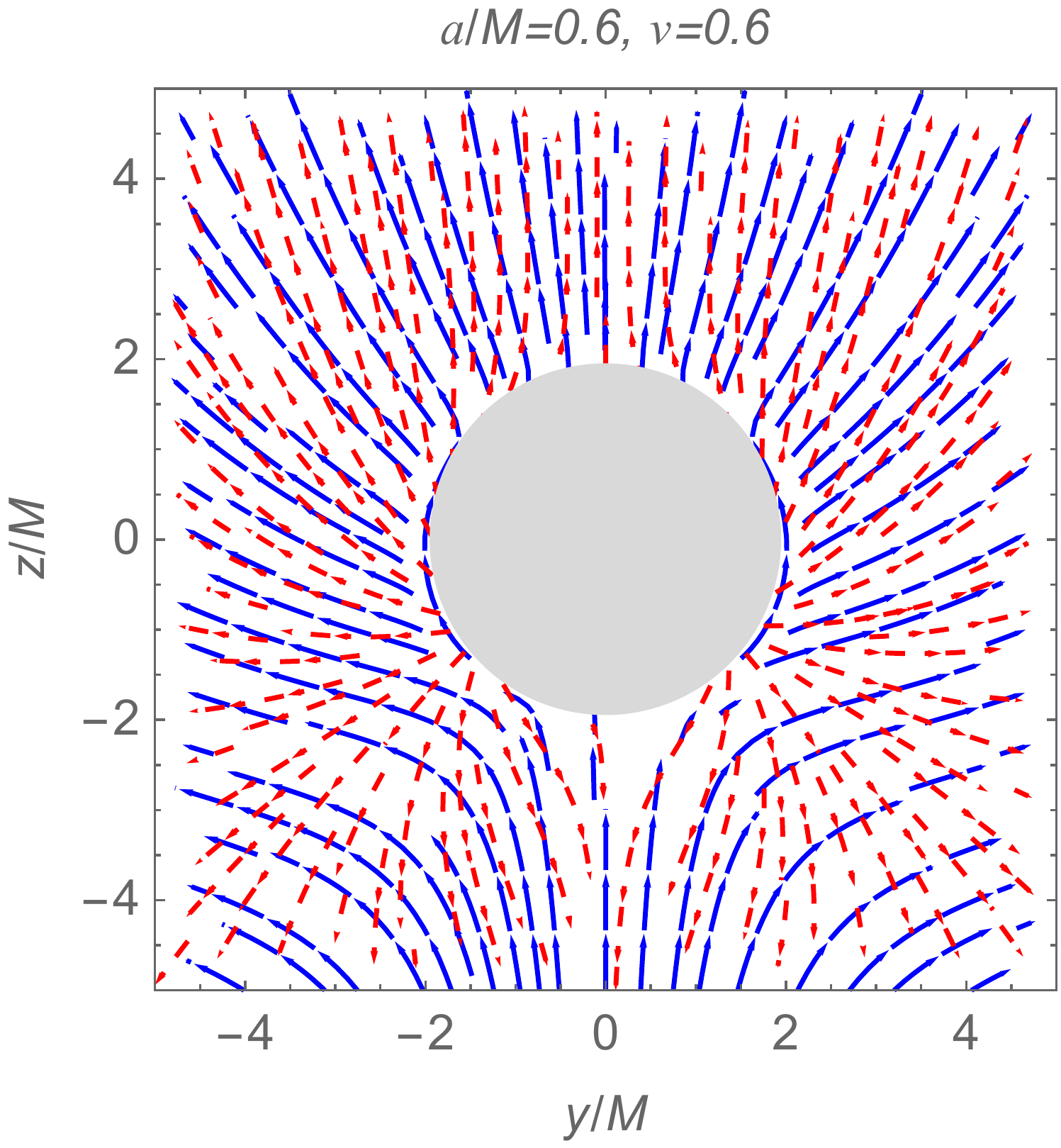}
\includegraphics[width=0.24\linewidth]{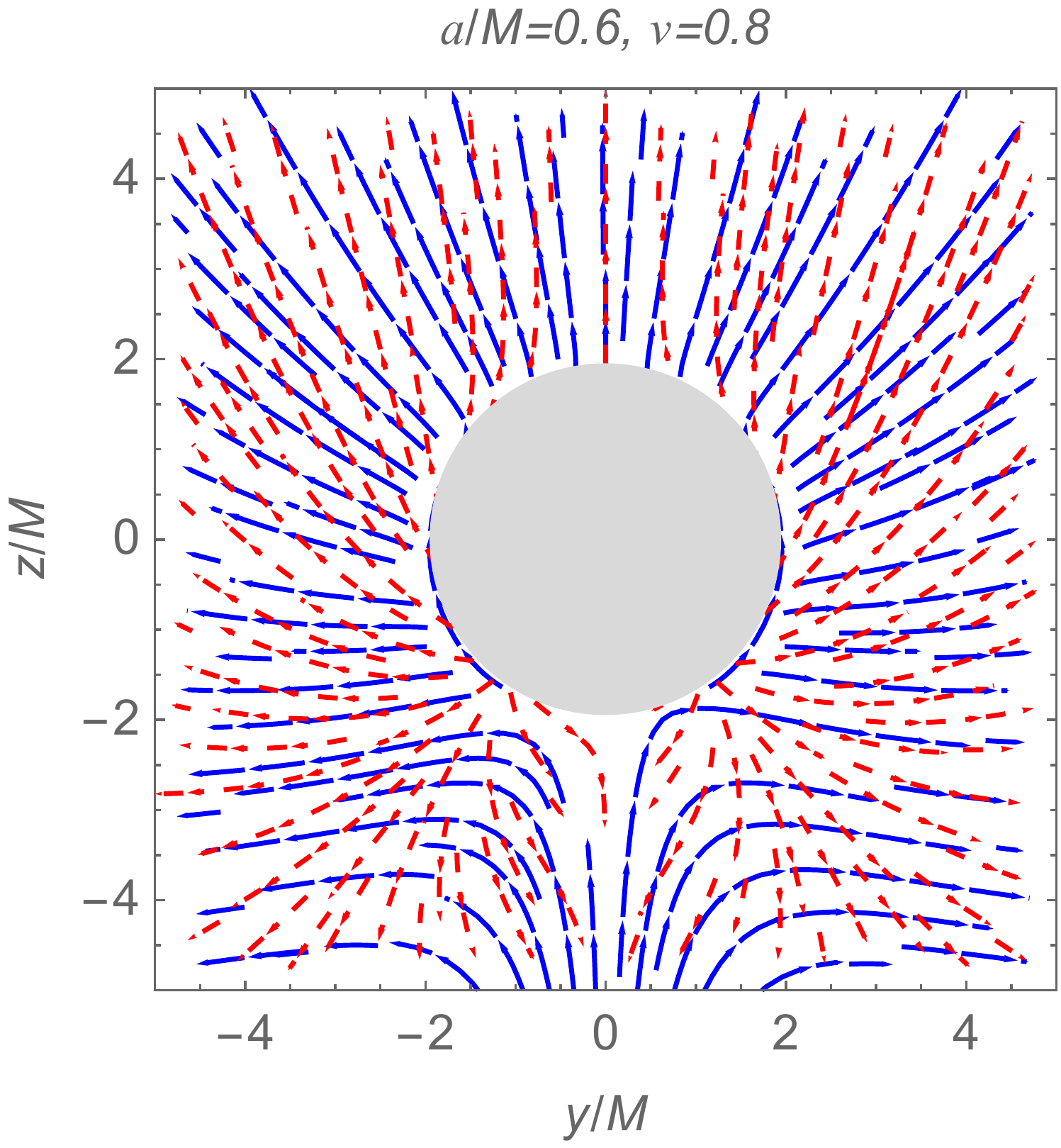}

\end{center}
\caption{The electric (red-dashed lines) and magnetic (blue solid lines) field lines for the different values of rotatin parameter $a$ and boosted velocity $v$ in the $y-z$ plane. The boost velocity is  aligned along the $z$-axis. \label{emfield}}
\end{figure*}

\begin{figure*}[t!]
\begin{center}

\includegraphics[width=0.31\linewidth]{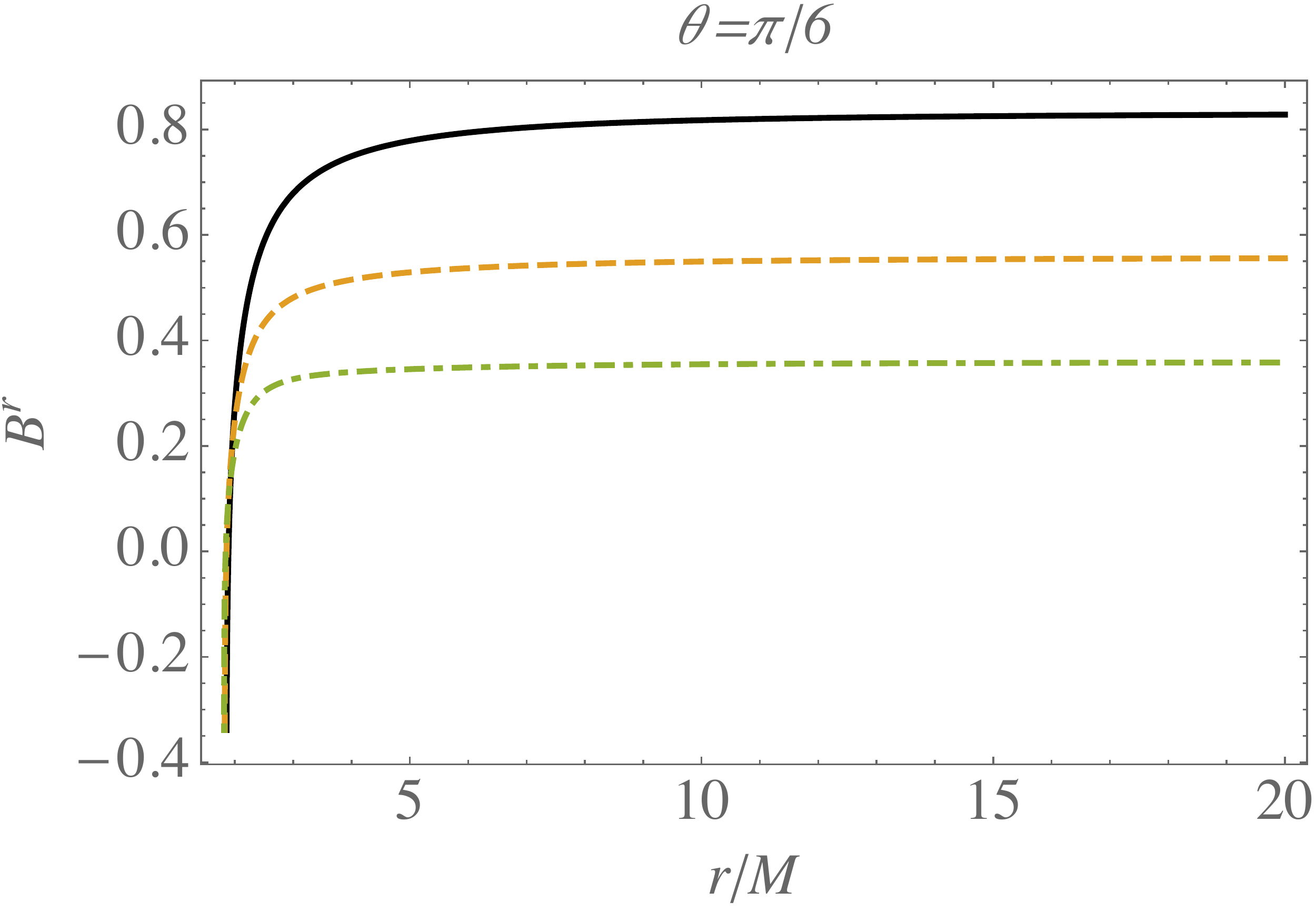}
\includegraphics[width=0.31\linewidth]{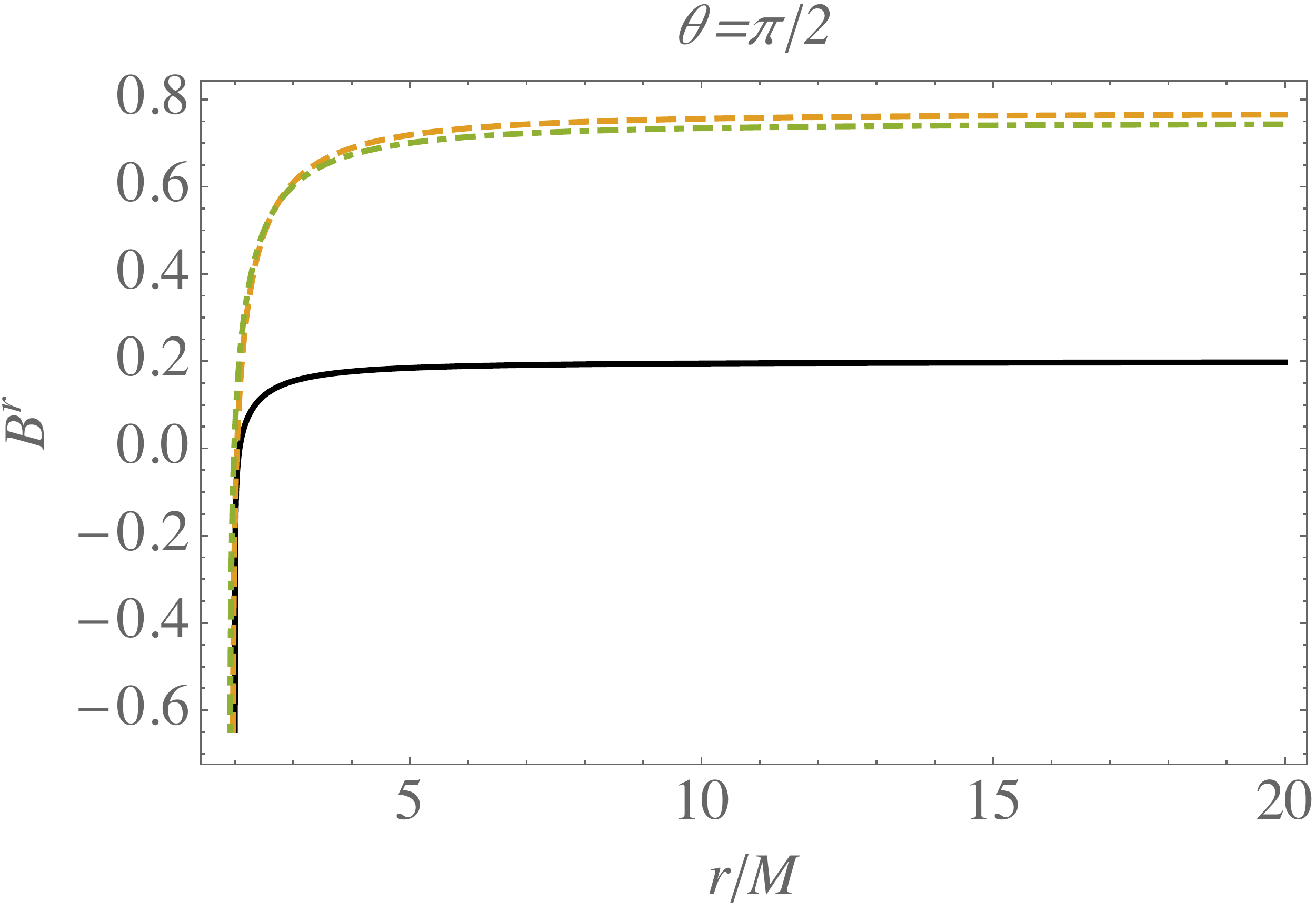}
\includegraphics[width=0.31\linewidth]{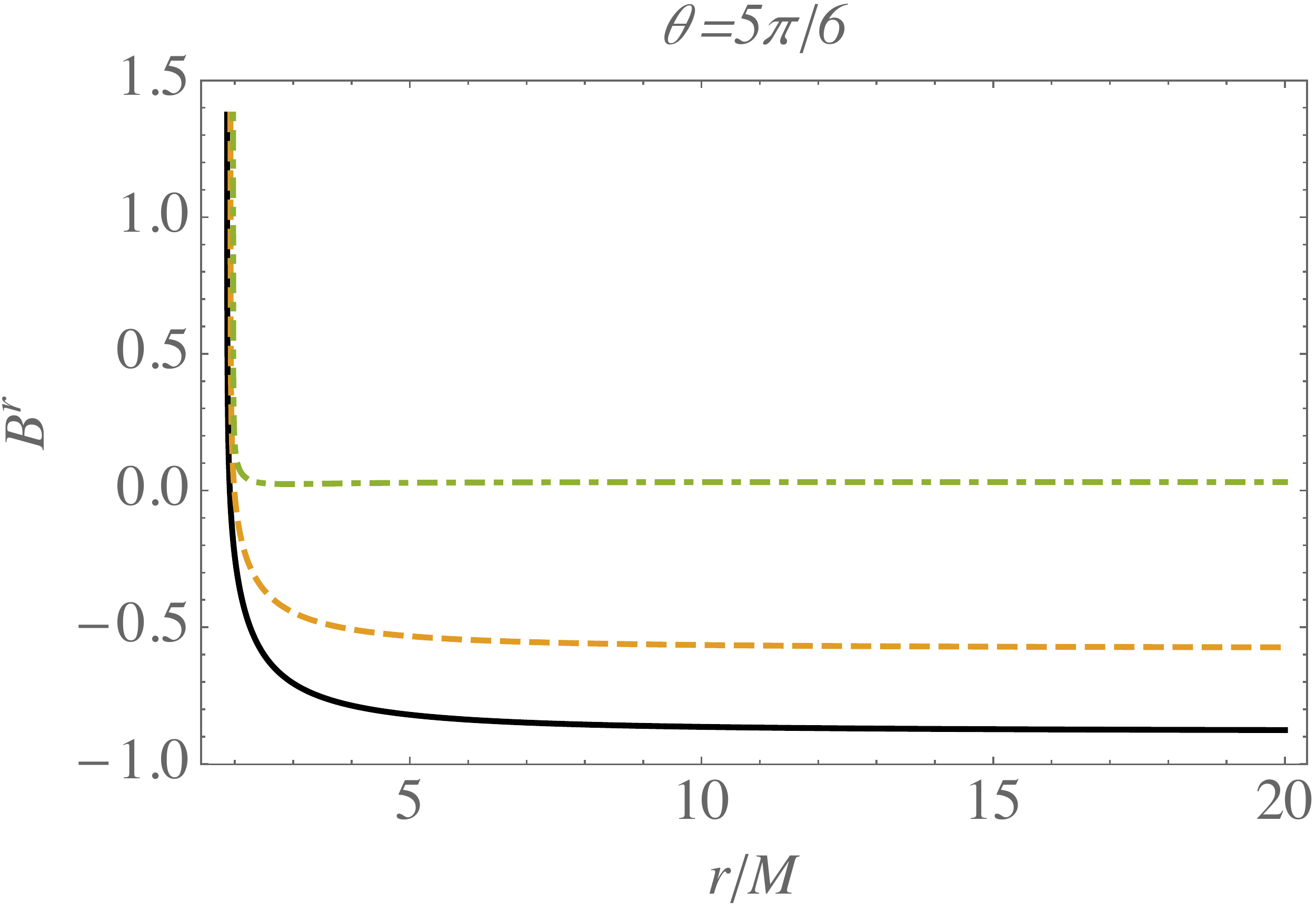}

\includegraphics[width=0.31\linewidth]{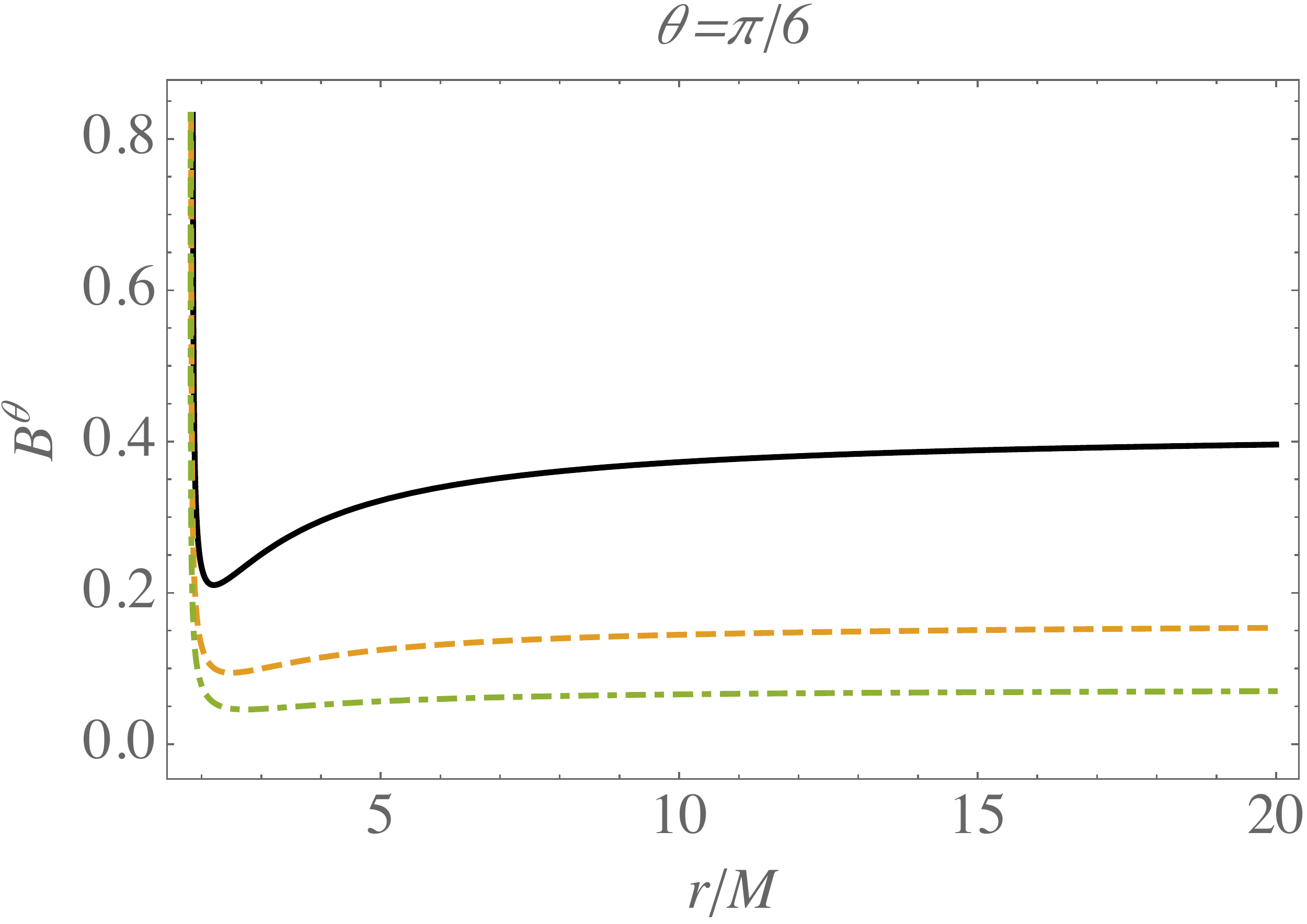}
\includegraphics[width=0.31\linewidth]{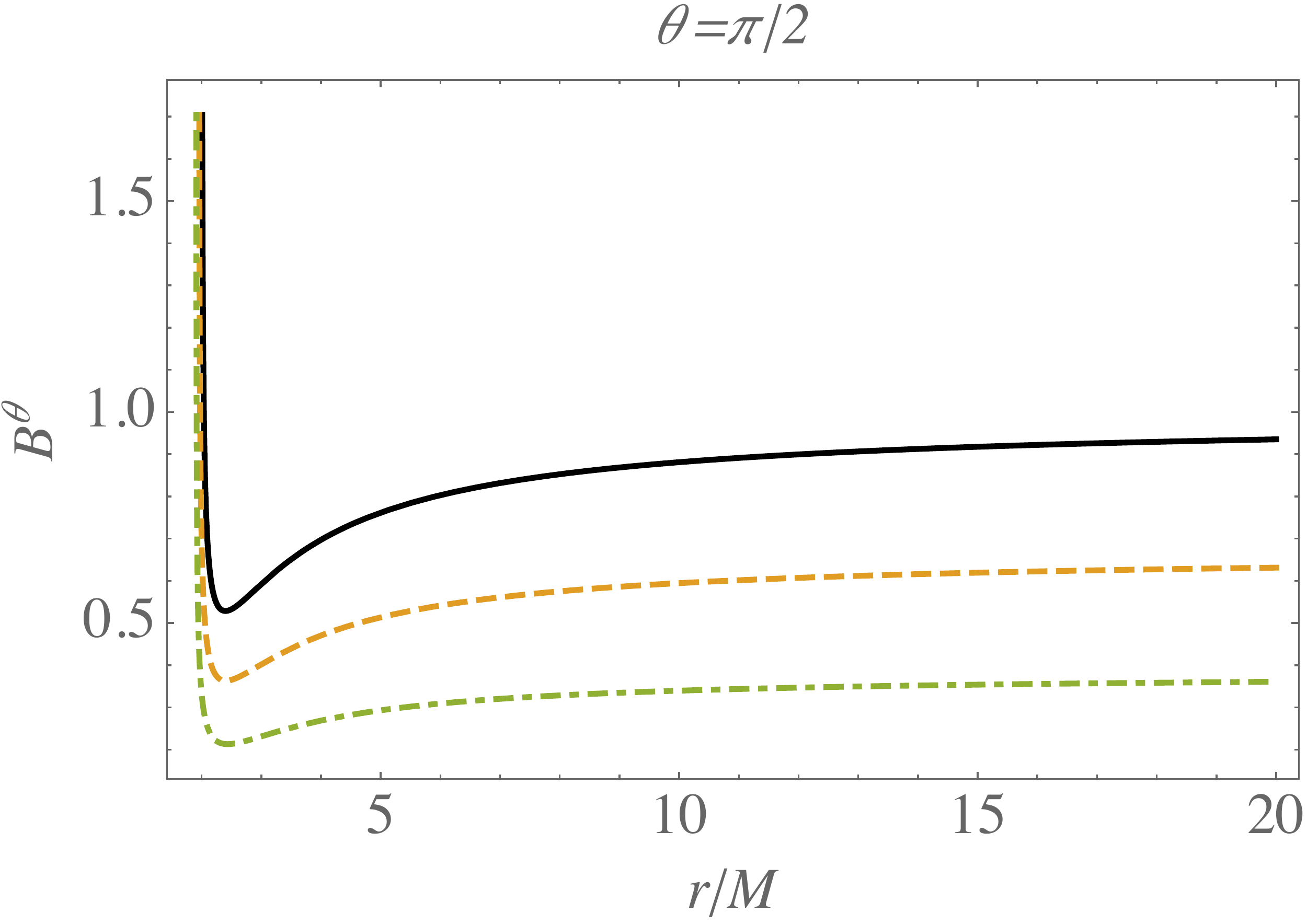}
\includegraphics[width=0.31\linewidth]{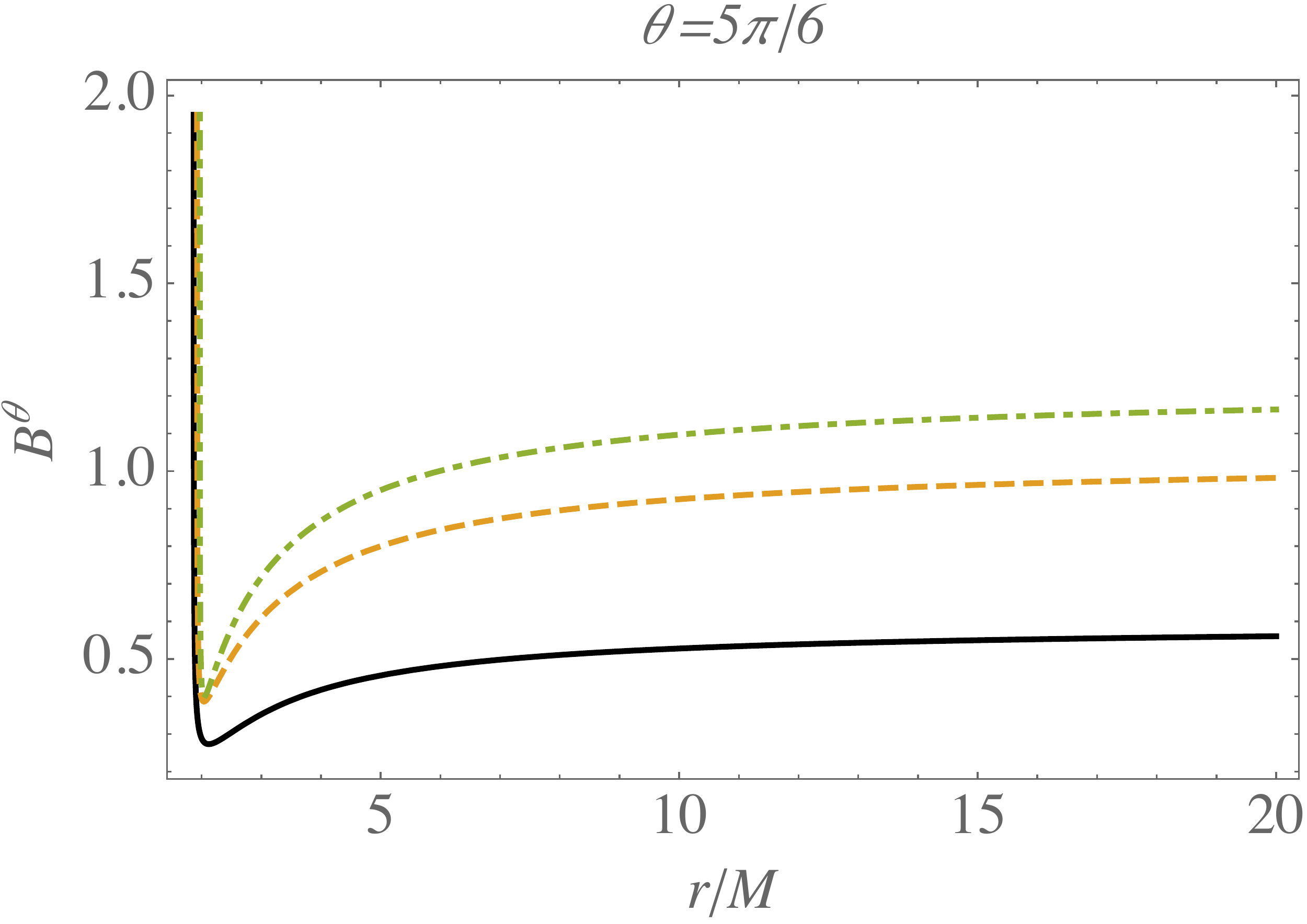}

\includegraphics[width=0.31\linewidth]{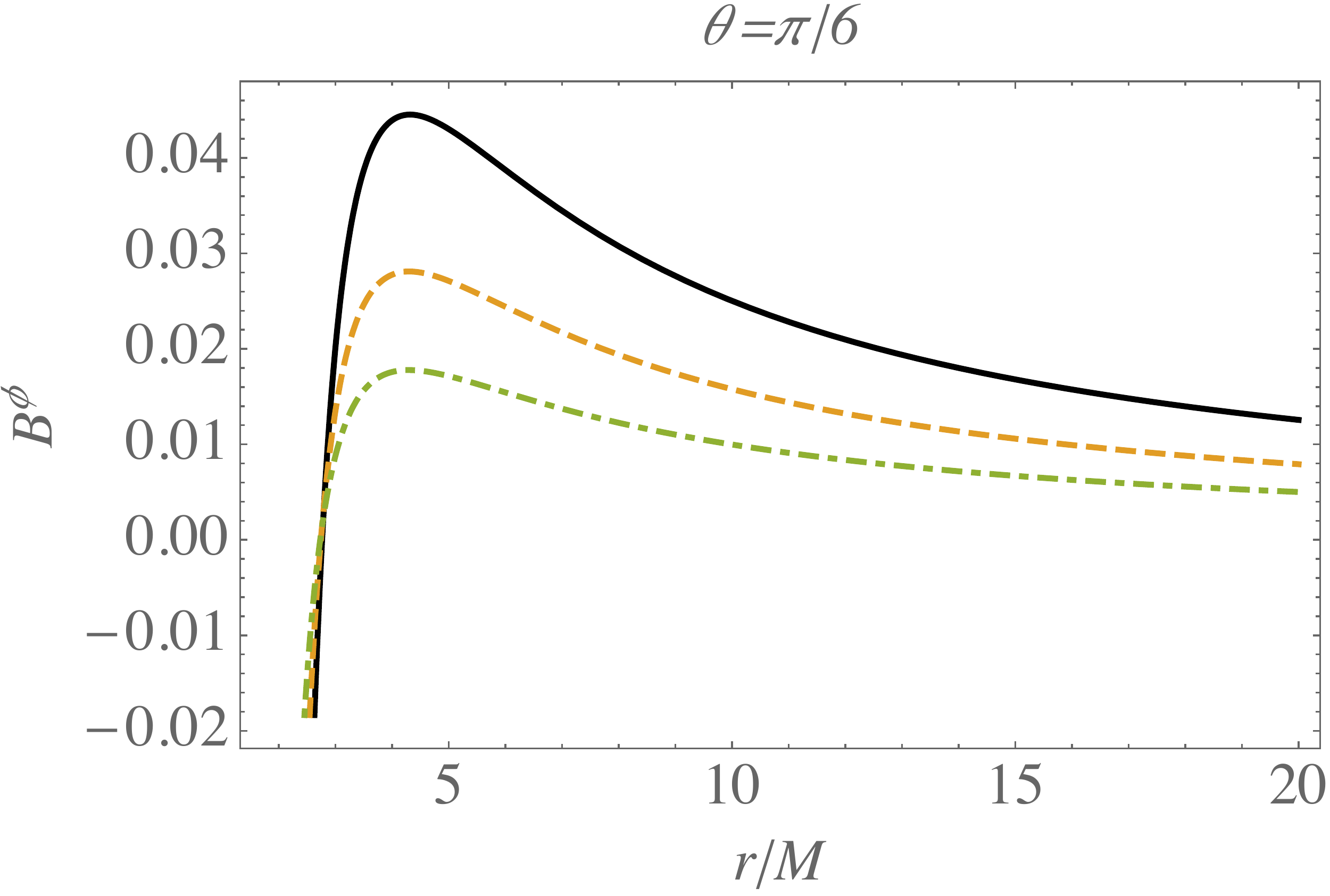}
\includegraphics[width=0.31\linewidth]{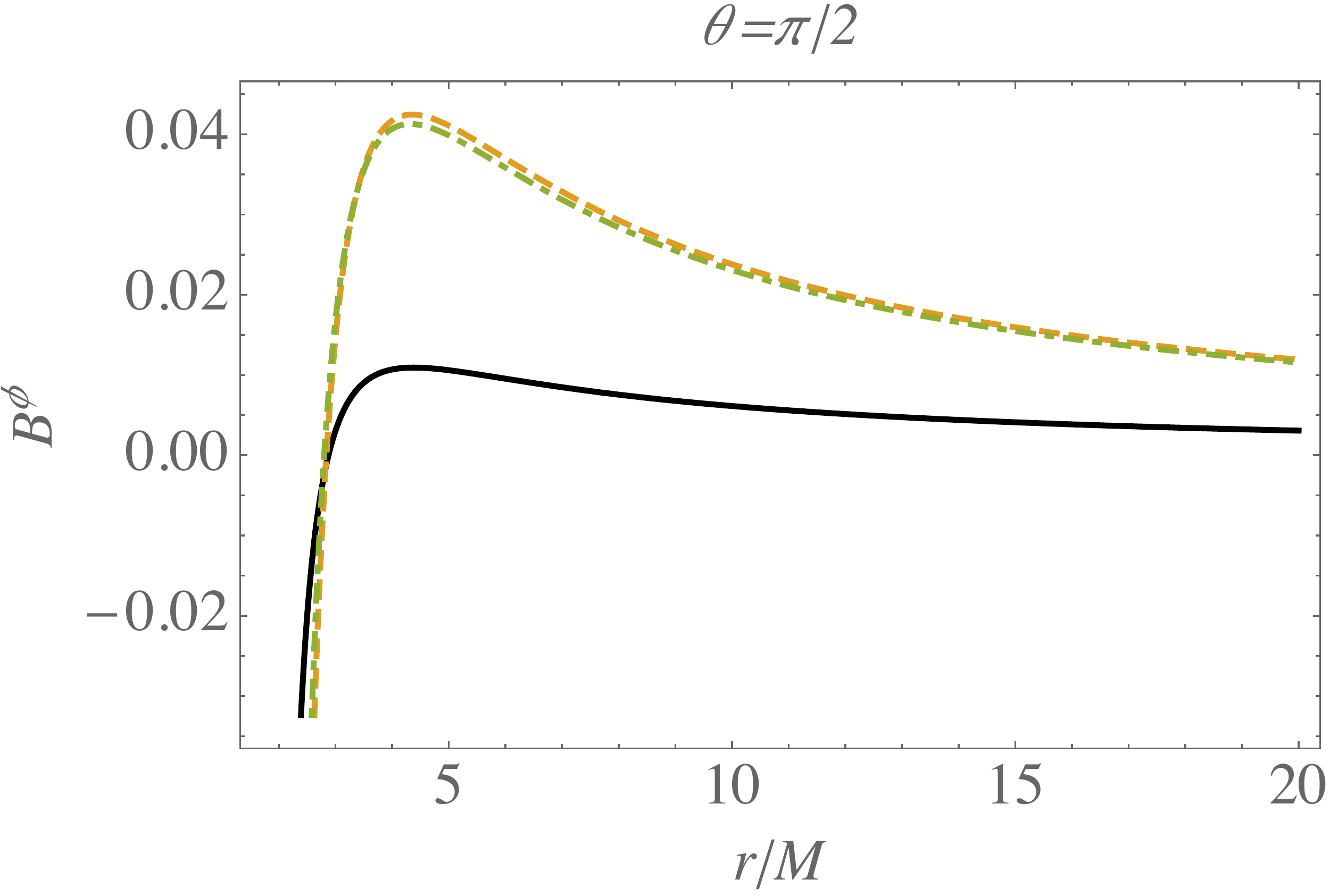}
\includegraphics[width=0.31\linewidth]{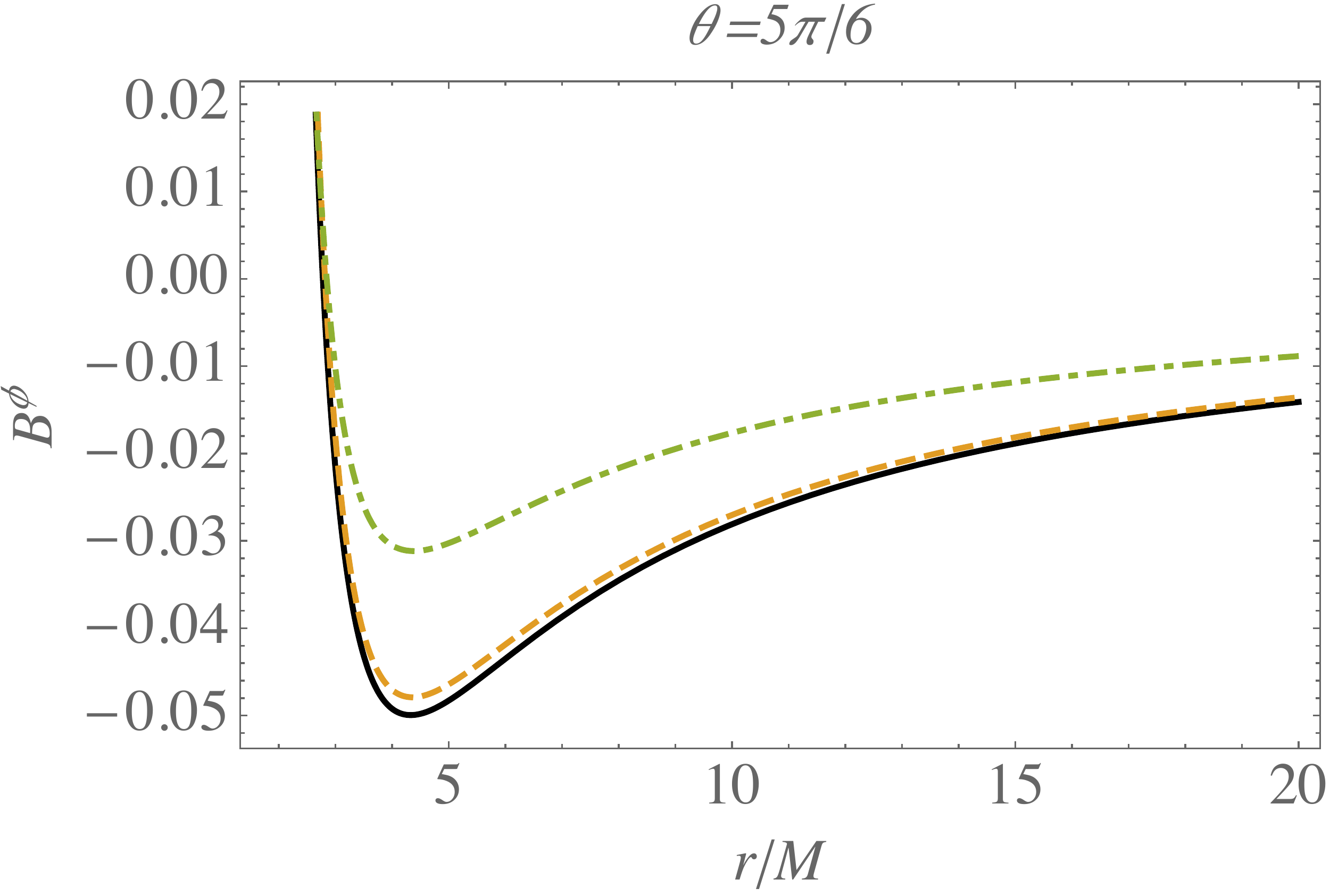}

\includegraphics[width=0.31\linewidth]{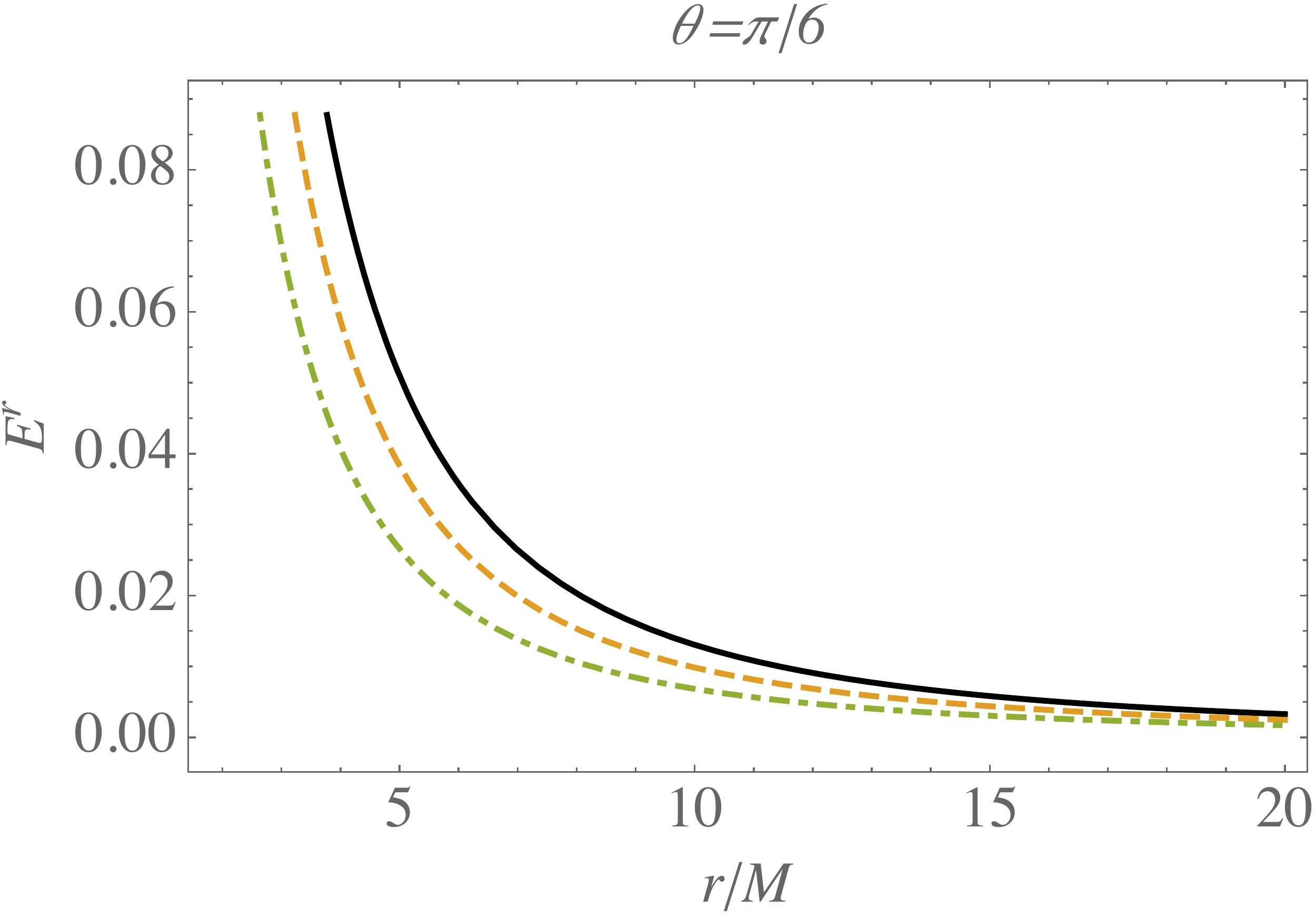}
\includegraphics[width=0.31\linewidth]{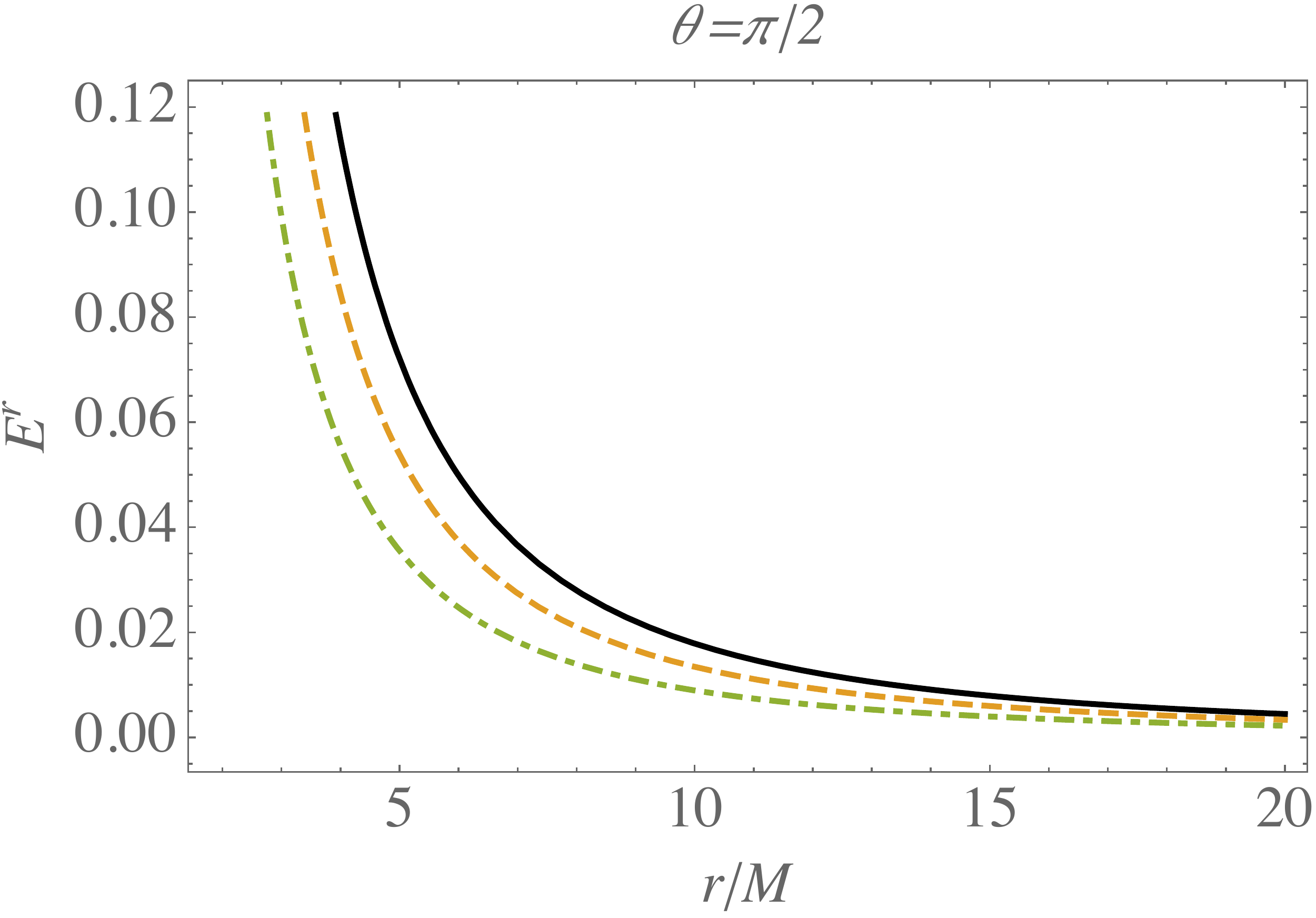}
\includegraphics[width=0.31\linewidth]{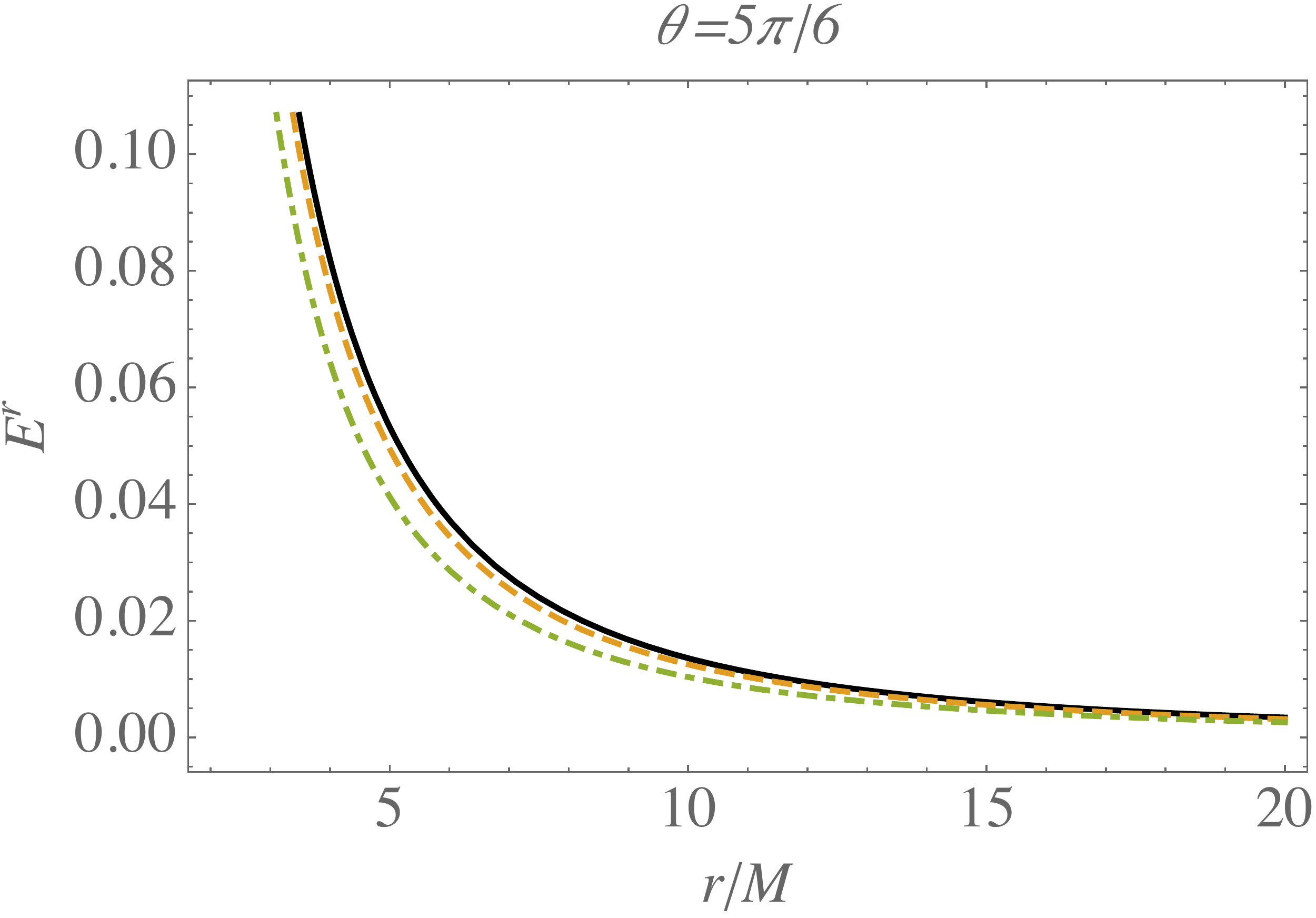}

\includegraphics[width=0.31\linewidth]{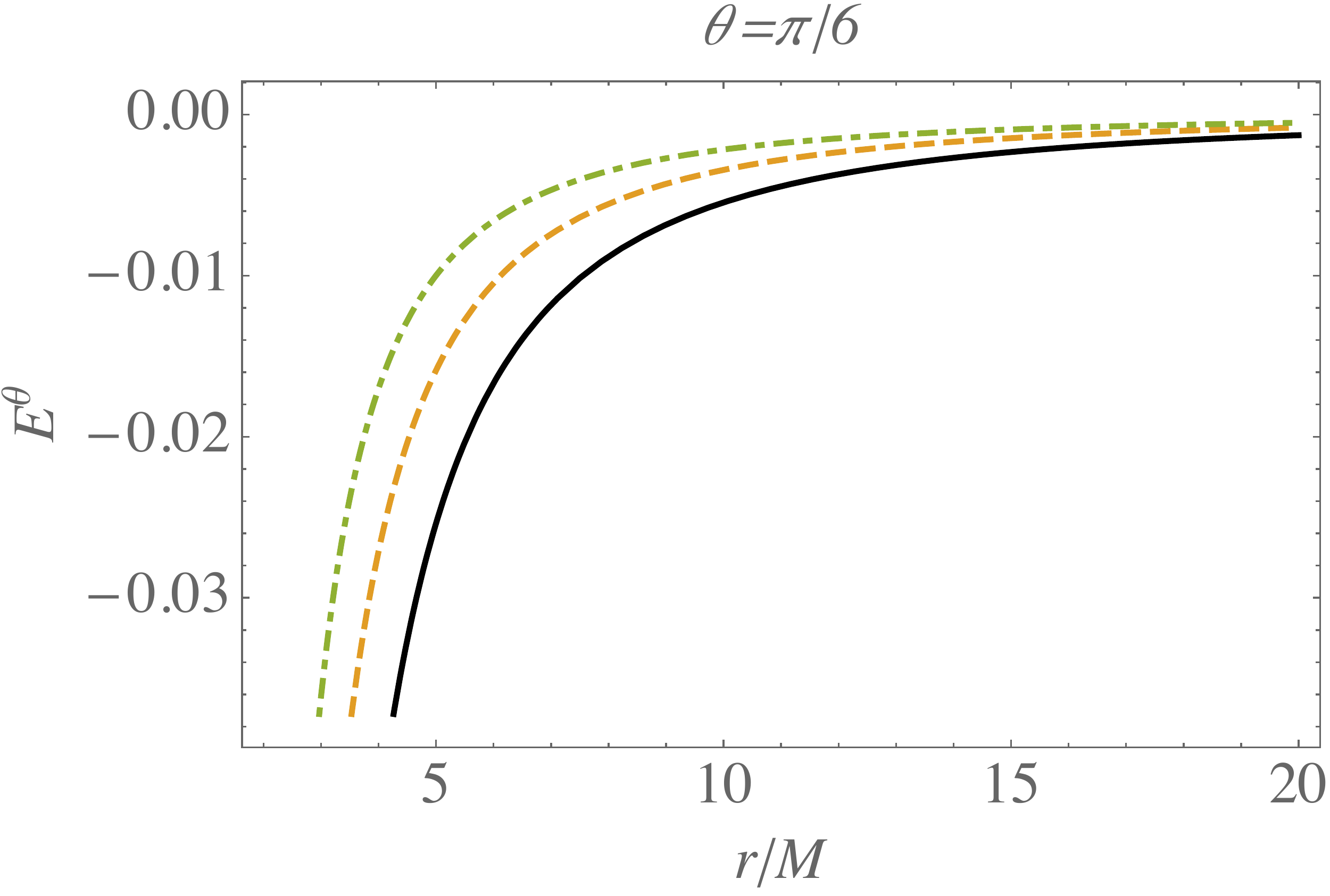}
\includegraphics[width=0.31\linewidth]{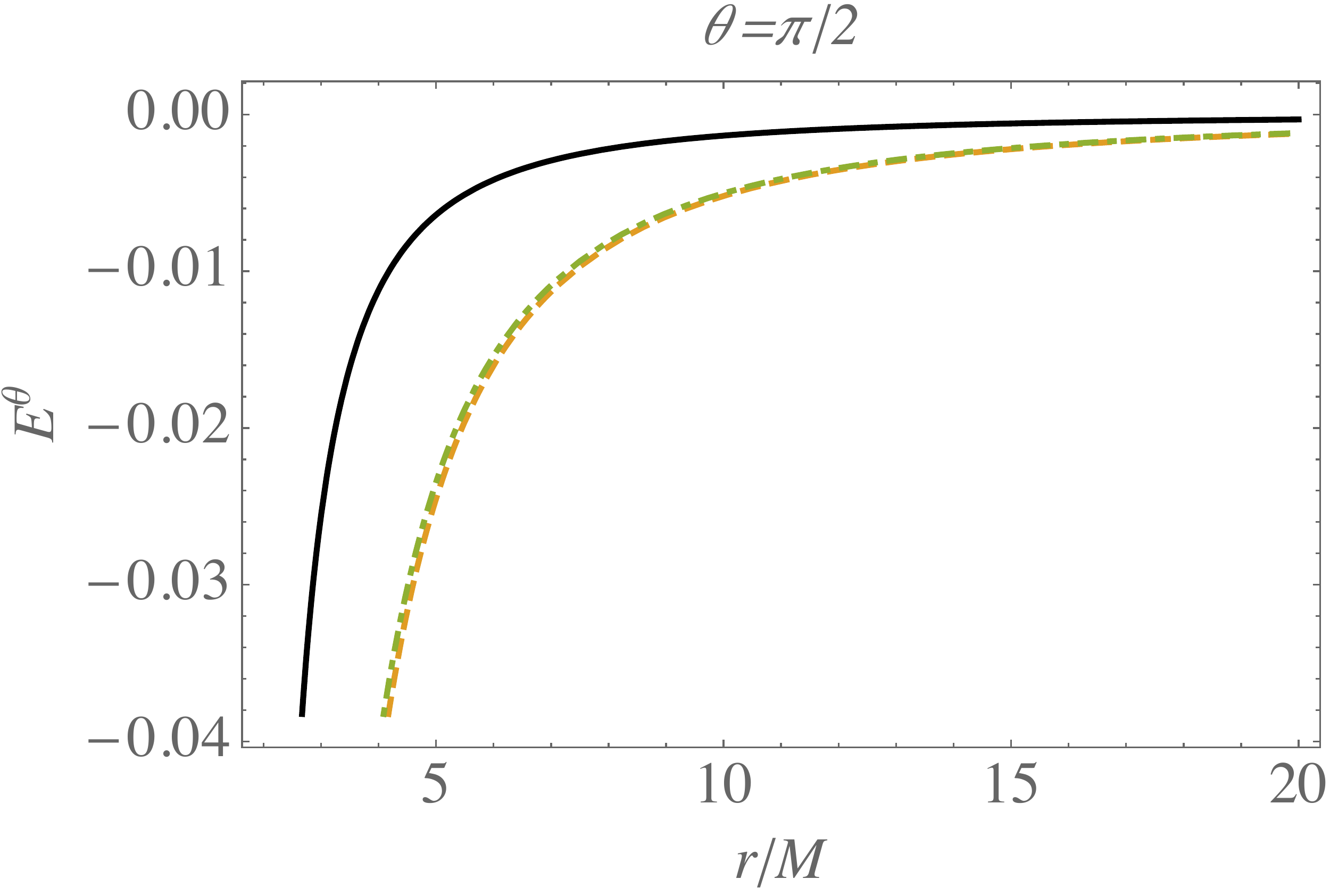}
\includegraphics[width=0.31\linewidth]{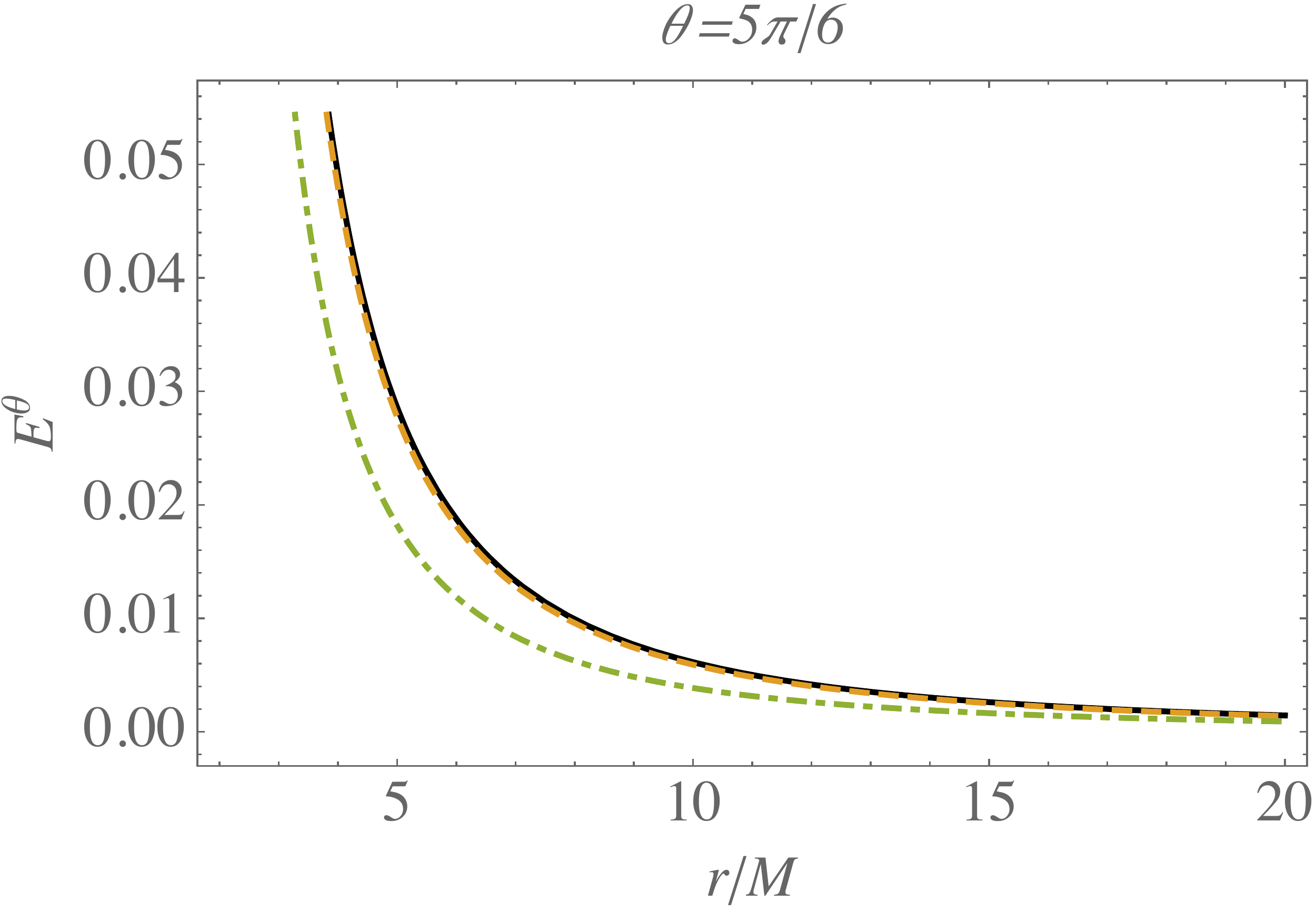}

\end{center}
\caption{The radial dependence of electric and magnetic field components for the different values of $v$: the solid, dashed and dot-dashed lines correspond to the values of $v=0.1; \  0.5;\ {\rm and}\ 0.9$, respectively.  The rotation parameter in all plots is taken to be $a/M=0.6$.  \label{eversusr1}}
\end{figure*}

\begin{figure*}[t!]
\begin{center}
\includegraphics[width=0.31\linewidth]{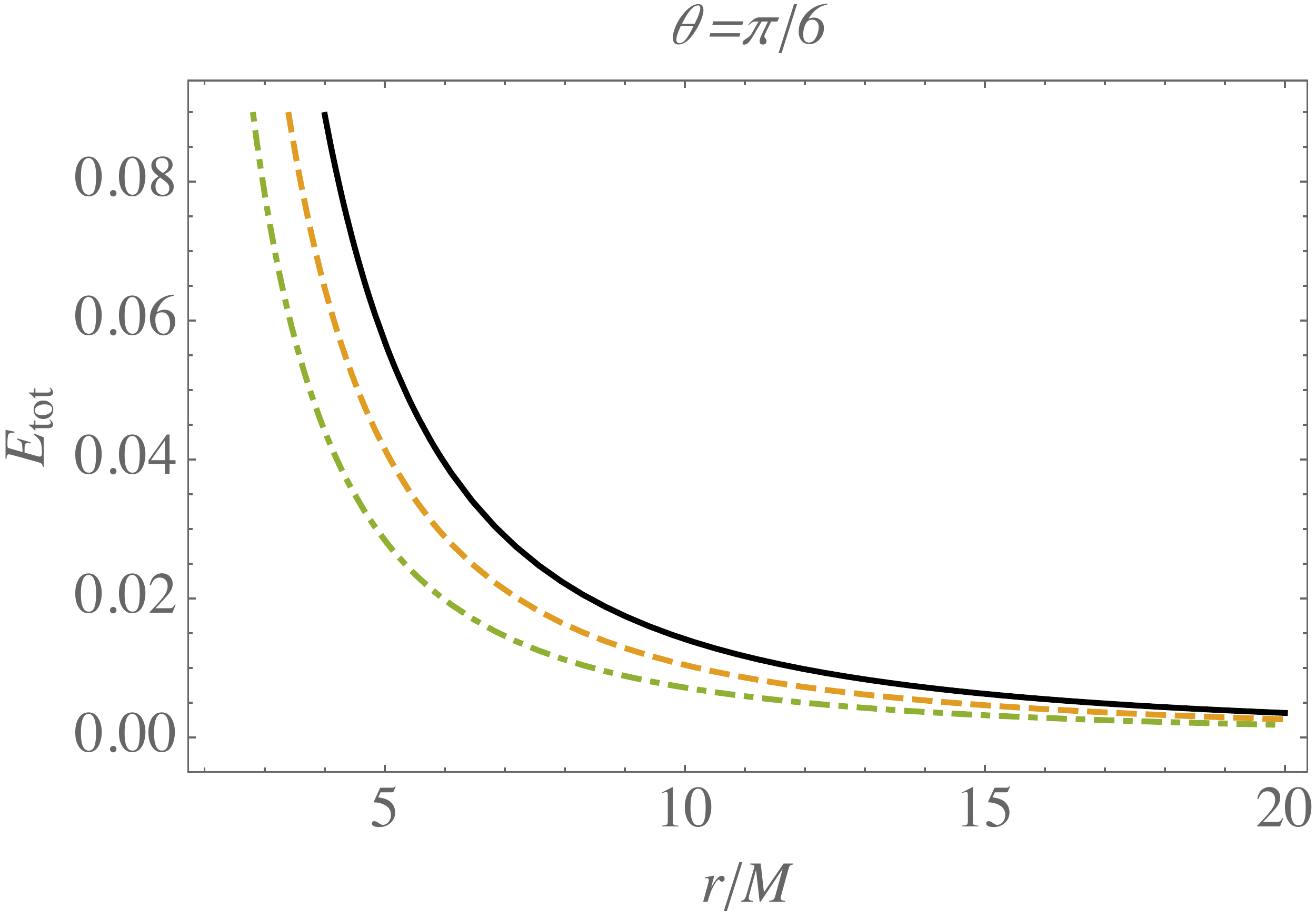}
\includegraphics[width=0.31\linewidth]{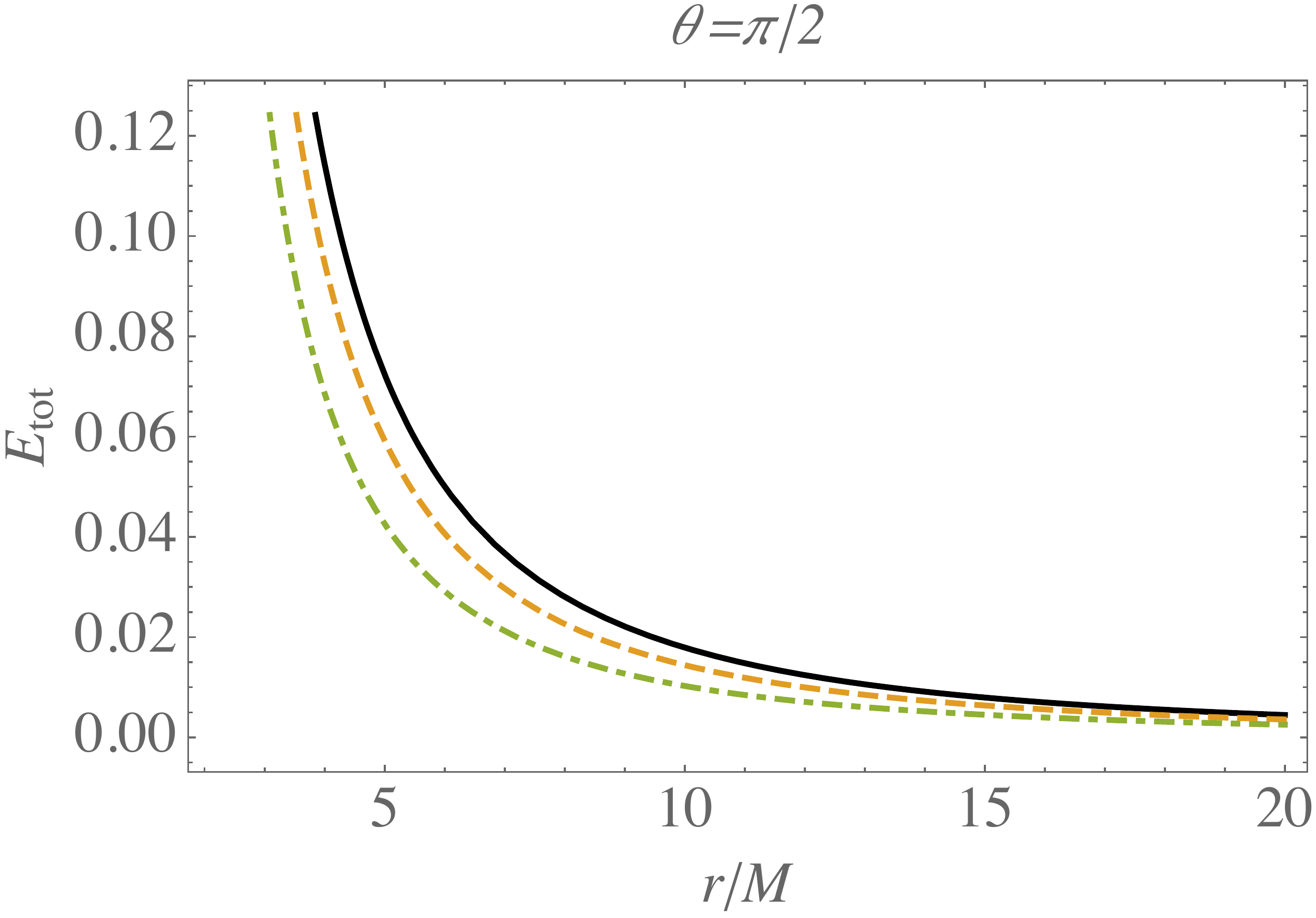}
\includegraphics[width=0.31\linewidth]{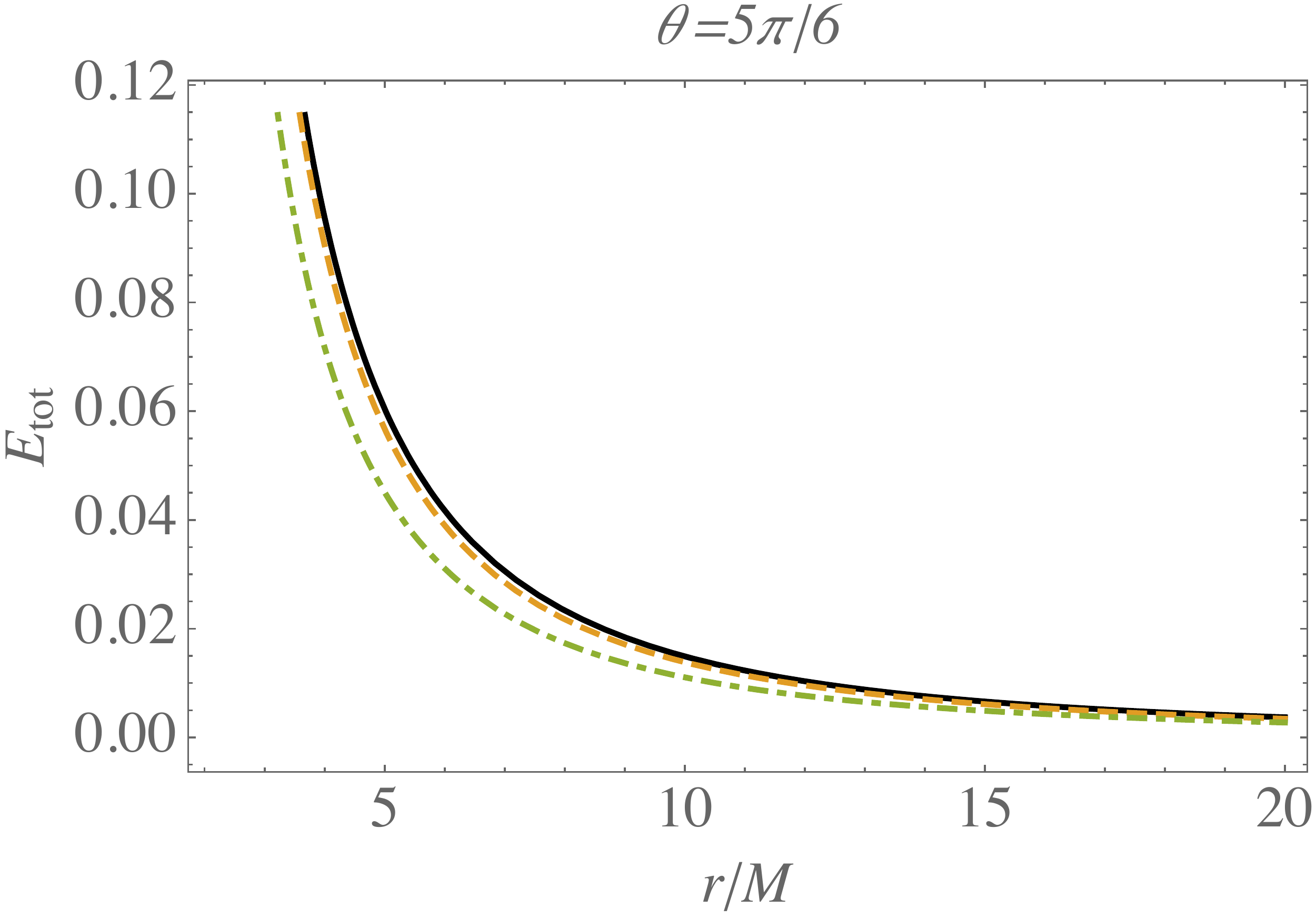}

\includegraphics[width=0.31\linewidth]{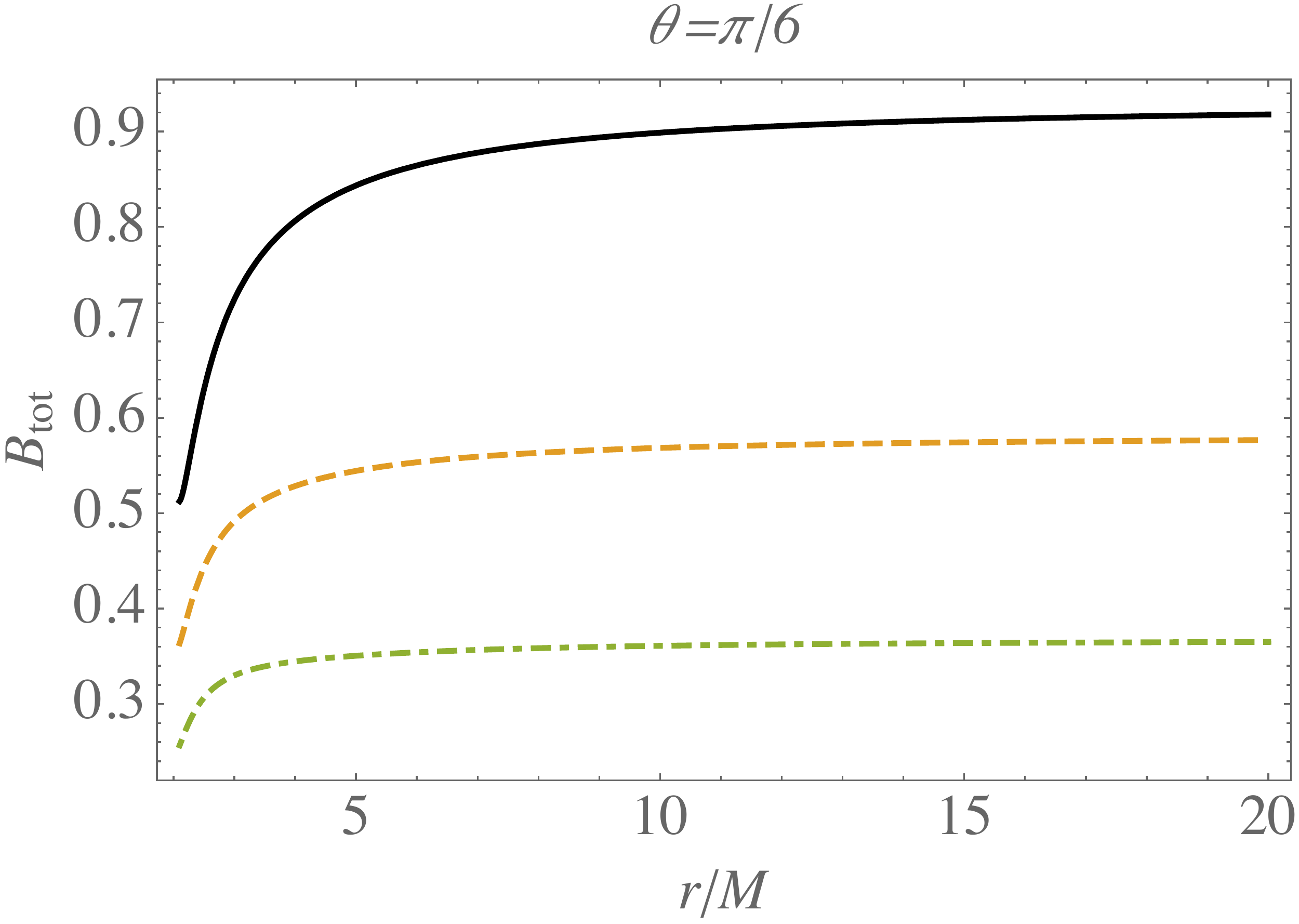}
\includegraphics[width=0.31\linewidth]{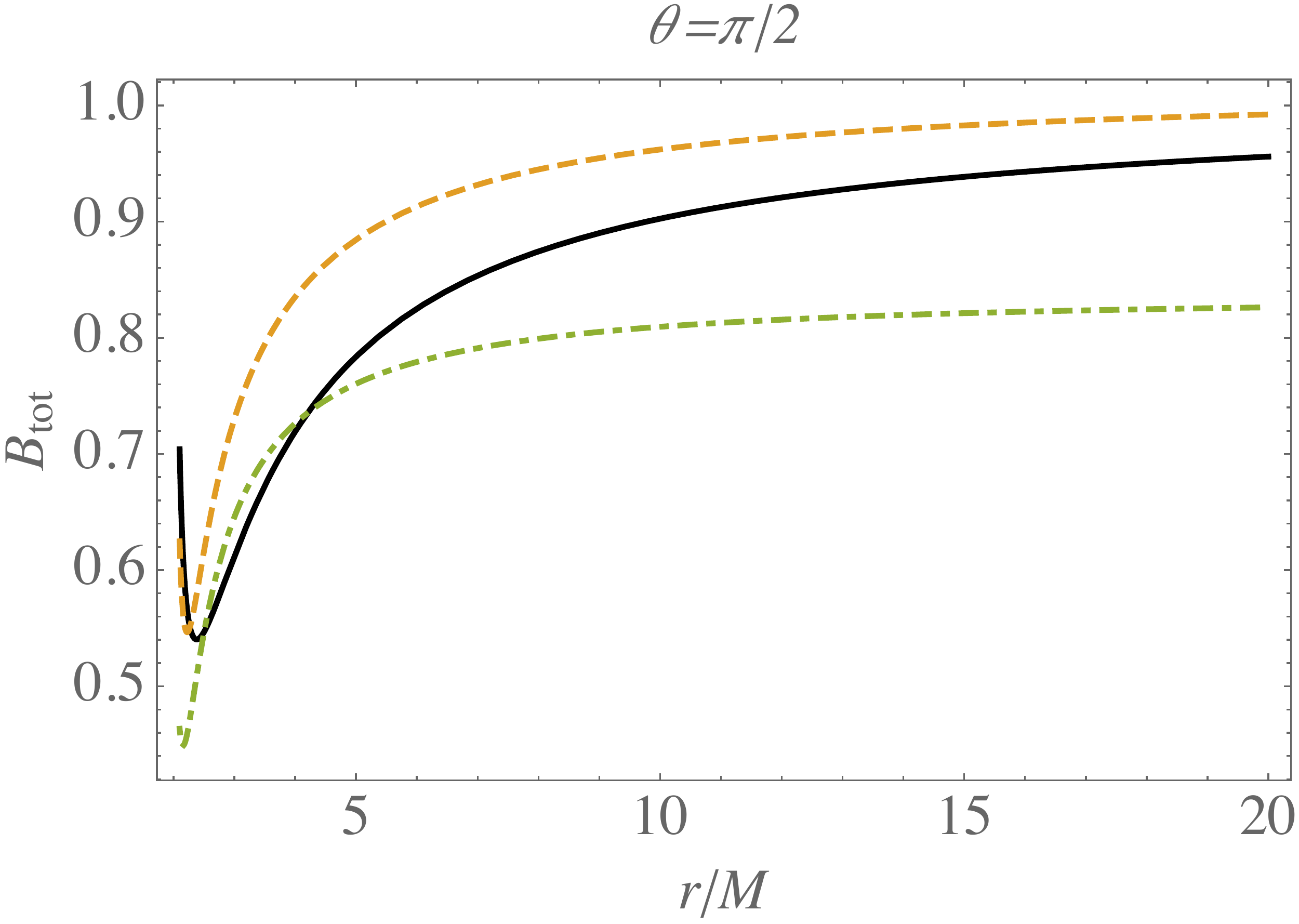}
\includegraphics[width=0.31\linewidth]{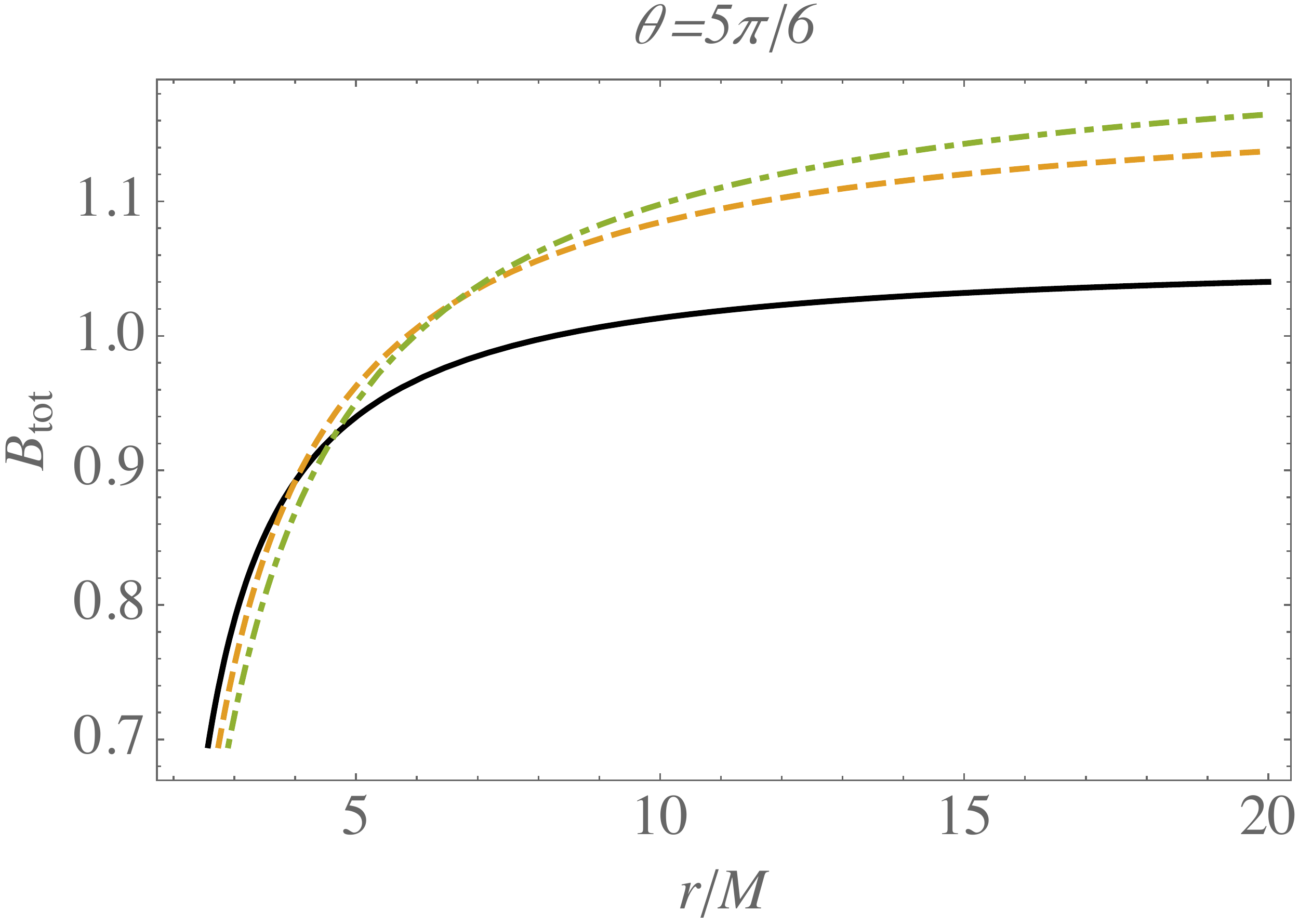}

\end{center}
\caption{The radial dependence of electric and magnetic field components for the different values of $v$: the solid, dashed and dot-dashed lines correspond to the values of $v=0.1; \  0.5;\ {\rm and}\ 0.9$, respectively.  The rotation parameter in all plots is taken to be $a/M=0.6$.     \label{bversusr1}}
\end{figure*}

%
%
%
%
%
%
%

In Fig.~\ref{emfield} the structure of the electromagnetic field for the different values of the rotation parameter and boost velocity is presented. Note that electric field is produced by twisting of magnetic field lines and hence its strength, as shown, increases with increase in rotation parameter. The effect of boost velocity is to sterngthen and weaken the field in forward and backward direction respectively.  To get a better understanding we plot in Figs.~\ref{eversusr1} and \ref{bversusr1} components of magnetic field as well as their total magnitude. As expected the the direction of the boost which is aligned to the rotation axis ($\theta = 0$ -- $z$ axis) is quite distinguished showing clear distinction between forward and backward direction. We shall next take up the study of charged particle trajectories.
	

\section{Charged particle motion \label{sect3}}

In this section we will study the charged particle motion with the rest mass $m$ and electric charge $e$ around boosted Kerr black hole in the presence of external magnetic field. Using the Hamilton-Jacobi equation
\begin{eqnarray}
g^{\mu \nu}\left(\frac{\partial S}{\partial x^\mu}-eA_\mu\right) \left(\frac{\partial S}{\partial x^\nu}-eA_\nu\right)=-m^2\ ,
\end{eqnarray}
one can find the equation of motion of the test particle where $S$ is the action of the particle in the background spacetime. The existence of the Killing vectors $\xi_{(t)}^\mu$ and $\xi_{(\phi)}^\mu$ allows us to write the action in the following form
\begin{eqnarray}
S=-{\cal E} t + {\cal L} \phi +S_{\rm r\theta}(r,\ \theta),
\end{eqnarray}

\begin{figure*}[t!]
\begin{center}
\includegraphics[width=0.32\linewidth]{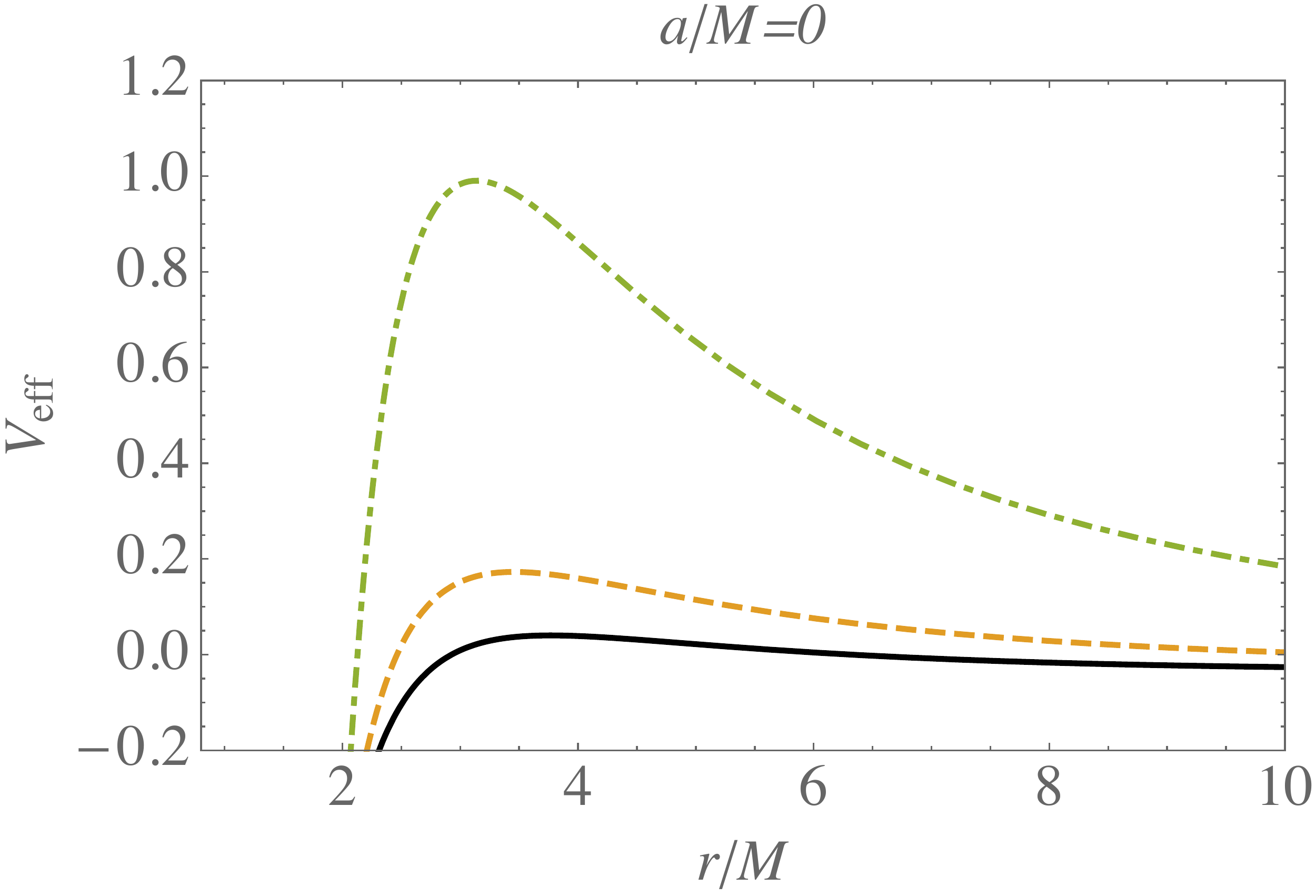}
\includegraphics[width=0.32\linewidth]{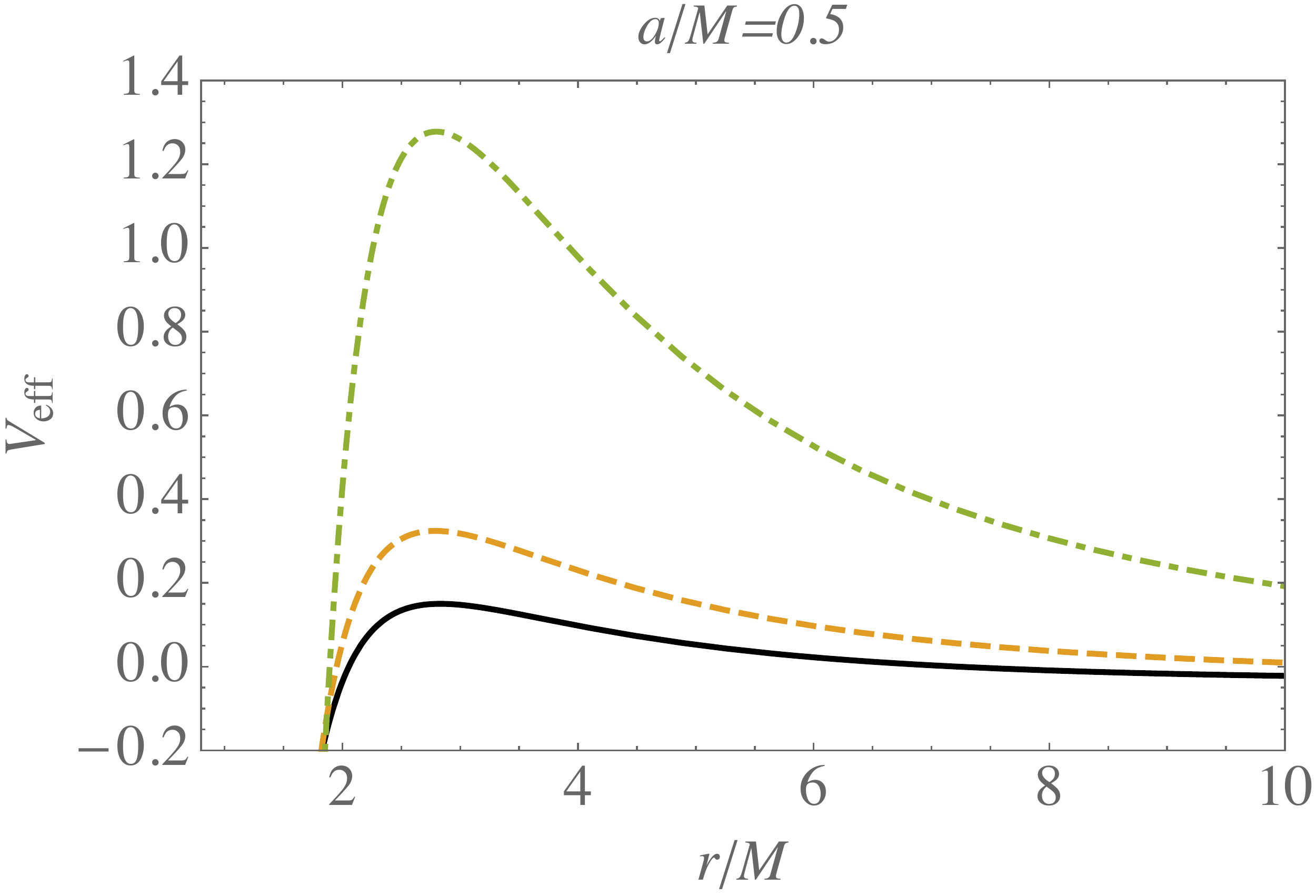}
\includegraphics[width=0.32\linewidth]{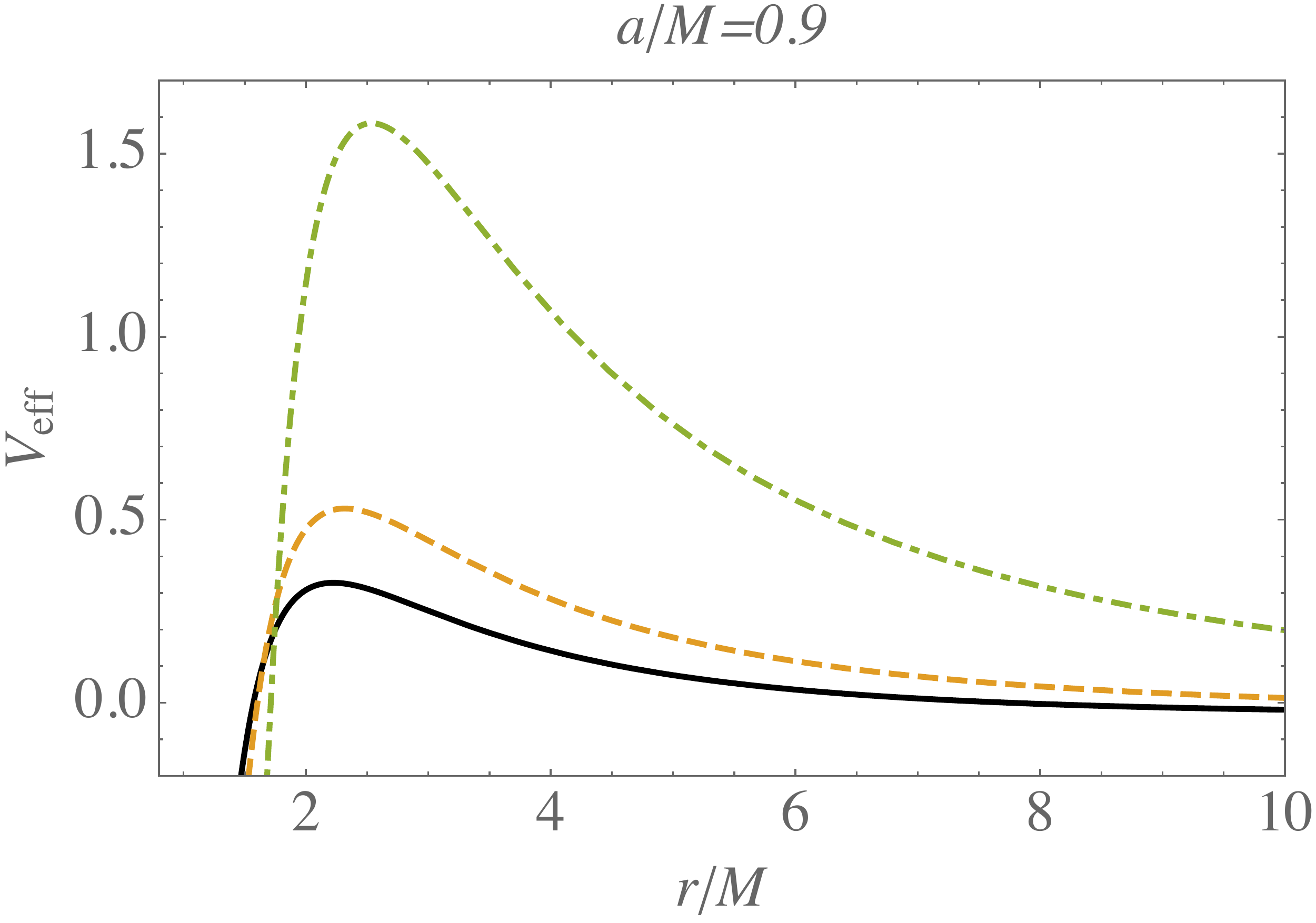}

\end{center}
\caption{The effective potential plots for different values of rotation parameter and the boost velocity: $v=0$ (solid line), $v=0.5$ (dashed line), $v=0.9$  (dotted line). \label{effpot1}}
\end{figure*}

First we will start with the case of either particle is neutral or there is no magnetic field;i.e. geodesic motion for a boosted Kerr geometry. Further we shall specialize to motion in
the equatorial plane ($\theta=\pi/2$) and write
\begin{eqnarray}
\Delta \frac{dt}{d\lambda}&=&a^2 {\cal E}^2 \left(1+\frac{2M}{r}\right) -2a {\cal L}\frac{M}{r} \gamma^2  \nonumber\\ &&+{\cal E} r^2 +2Mr\sqrt{\cal R}\\
\frac{\Delta}{\gamma^2} \frac{d\phi}{d\lambda}&=& {\cal L} \left(1-\frac{2M}{r}\right) \gamma^2 +\frac{2 a{\cal E} M}{r} +a 	\sqrt{\cal R}\ ,
\\
\left(\frac{dr}{ds}\right)^2&=& {\cal R}(r) ={\cal E}^2-1-2 V_{\rm eff} = \nonumber \\&& {\cal E}^2 -1 +\frac{2M}{r}
+\frac{a^2}{r^2} \left[{\cal E}^2 \left(1+\frac{2M}{r}\right)-1\right] \nonumber \\&& - \frac{4a{\cal E}{\cal L}M \gamma^2}{r^3} -\frac{{\cal L}^2 \gamma^4}{r^2}\left(1-\frac{2M}{r}\right) \label{reqnulmf}
\end{eqnarray}
where $\Delta=r^2-2 Mr+a^2$.  The effective potential, $ V_{\rm eff}$ for radial motion is plotted for various values of boost velocity as shown in Fig.~\ref{effpot1}. It shows that maximum of the potential increases with increase in rotation and boost velocity. It is well known that rotation weakens gravitational field -- gravitational attraction closer to a rotating black hole is weaker than that of the Schwarzschild one. Similarly boost velocity also works in consonance with the rotation parameter and thereby enhancing weakening effect.

\begin{figure*}[t!]
\begin{center}
\includegraphics[width=0.32\linewidth]{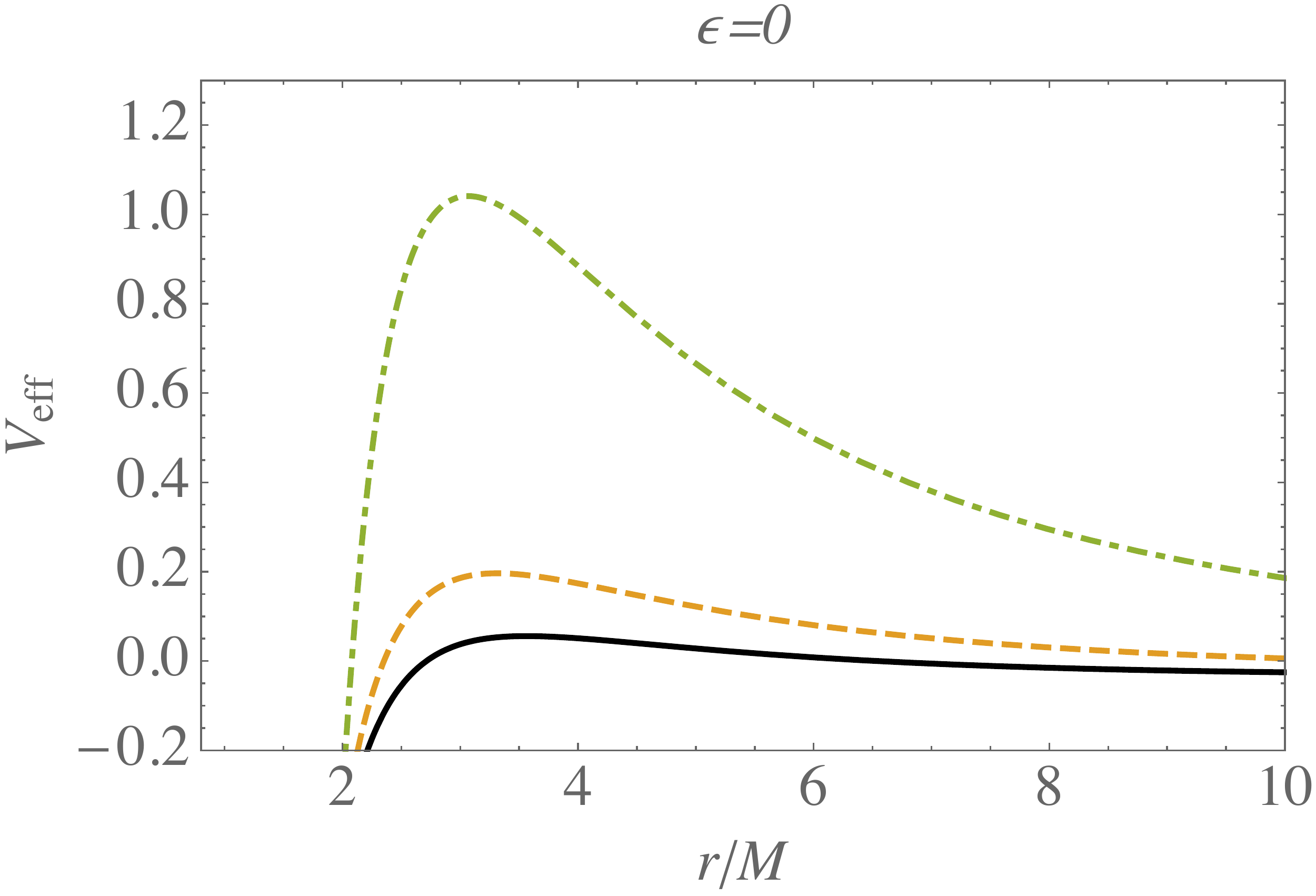}
\includegraphics[width=0.32\linewidth]{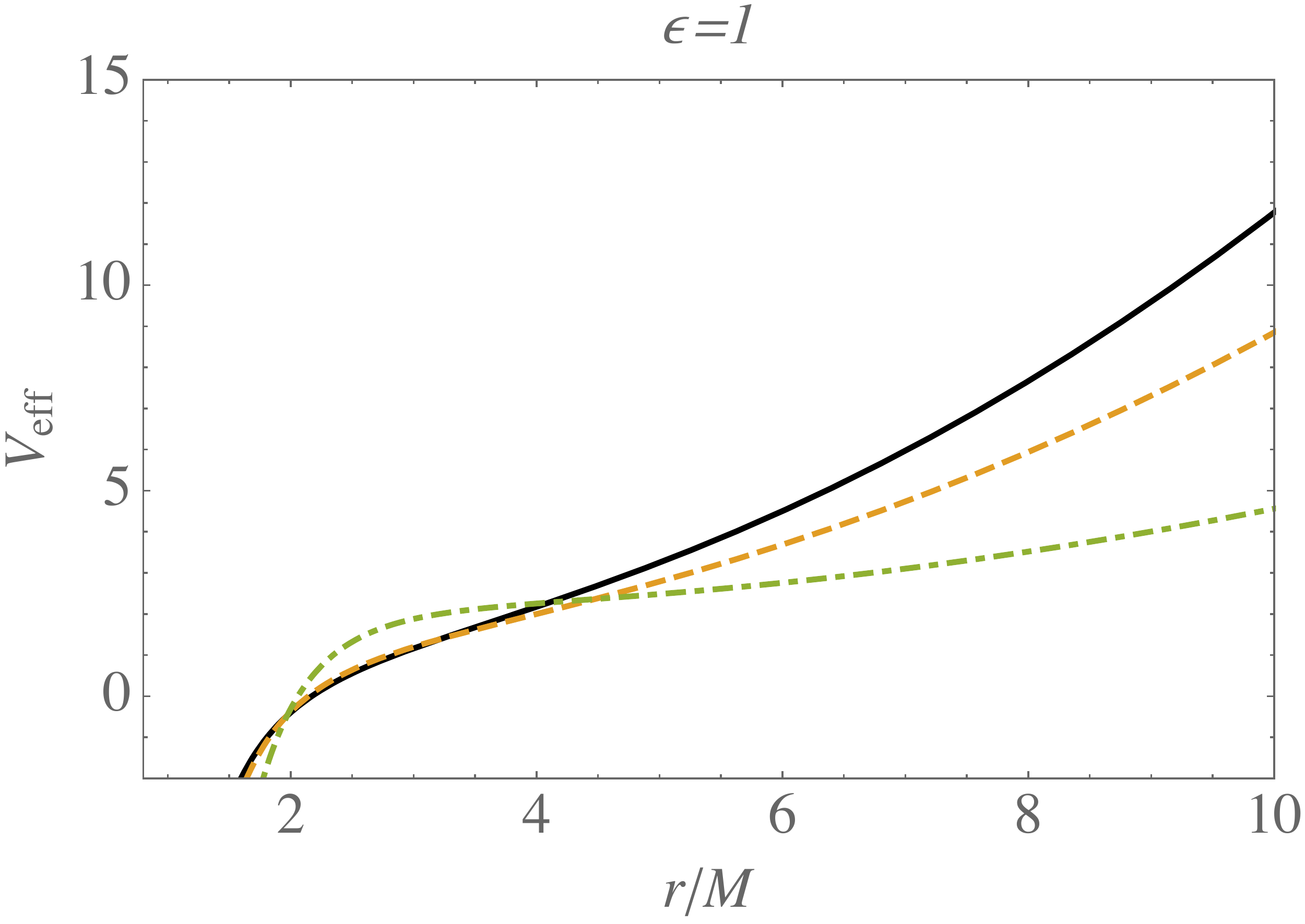}
\includegraphics[width=0.32\linewidth]{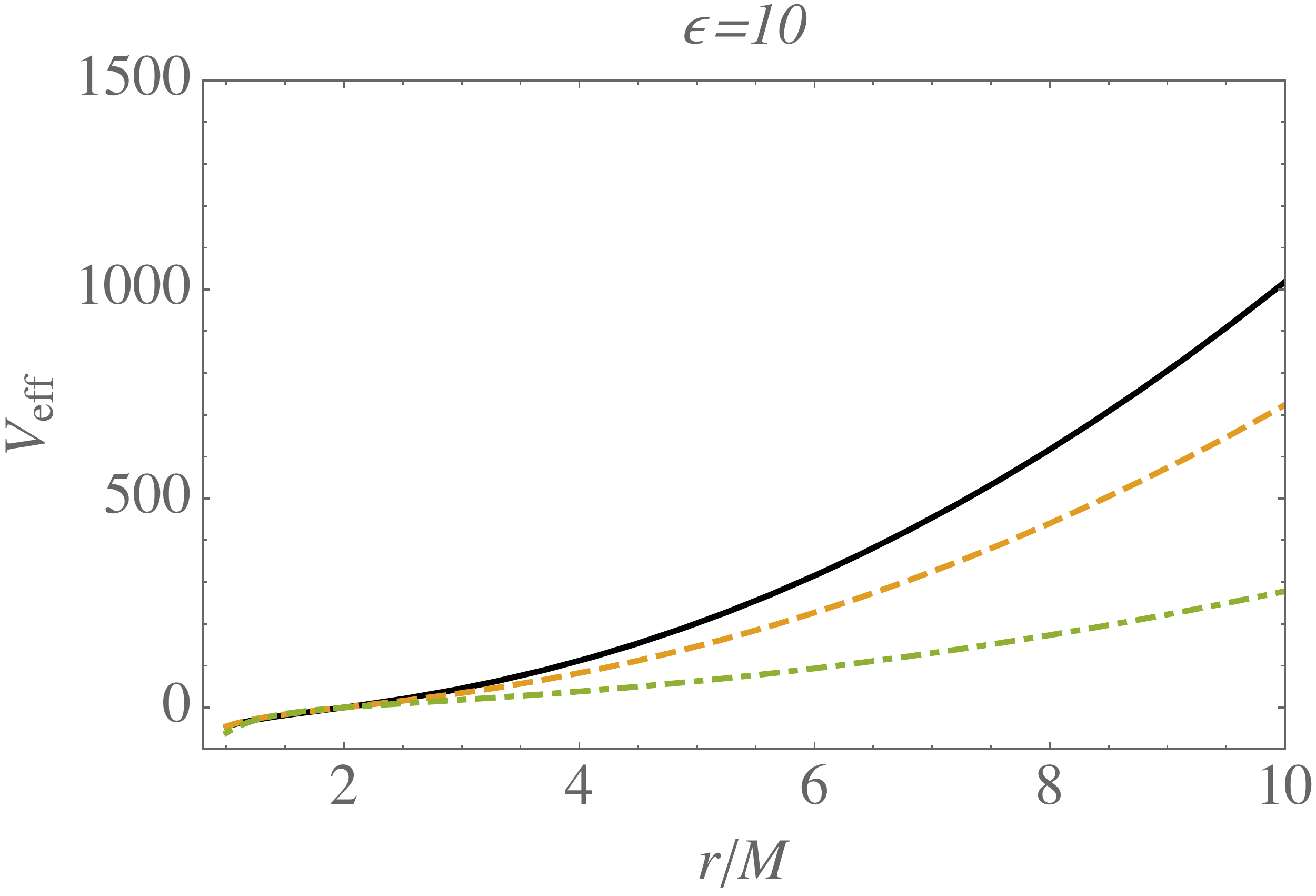}

\end{center}
\caption{The radial dependence of the effective potential of the radial motion of the charged particle for the different values of magnetic parameter $\epsilon$ parameter and the boost velocity: $v=0$ (solid line), $v=0.5$ (dashed line), $v=0.9$  (dotted line). \label{effpot2}}
\end{figure*}

Now we switch on the magnetic field, then the radial function $\cal R$ will be modified as
\begin{eqnarray}
&& {\cal R}(r) = \nonumber \\
&&\frac{1}{4 r^4}\Bigg[4 (a \gamma^2 {\cal L} - 2 {\cal E} M r)^2 - \frac{\Delta}{\gamma^4}\bigg\{4 \gamma^8 {\cal L}^2 - 8 a^2 \gamma^2 \epsilon^2 k M r \nonumber\\&& -
    4 \gamma^4 r (2  M {\cal E}^2 - 2  M a^2 \epsilon^2 k^2+ ({\cal E} + a \epsilon k)^2 r-r -
           {\cal L} \epsilon r ) \nonumber\\&& +
     \epsilon^2 r^4 +  2 M a^2 \epsilon^2 r +a^2 \epsilon^2 r^2\bigg\}\Bigg]
\end{eqnarray}
where $\epsilon=eBM/m$ is the parameter responsible for the electric charge to electromagnetic field interaction. The effective potential plots in Fig.~\ref{effpot2} shows that near  the horizon, it is the boost velcoity that domeinates while as $r$ increases electromagnetic field interaction has the dominant effect. This point is very important in the context of radius of innermost stable circular orbits (ISCO)  which determine the inner edge of accretion disk. This is what we take up next.

As the next step we consider the innermost stable circular orbits (ISCO). For this we will use the equations ${\cal R}=0=d{\cal R}/dr$ with the condition $d^2 {\cal R}\leq 0$ where the former two are the conditions for existence of circular orbit while the latter is the stability condition. Withoult loosing the generality one may put  $M=1$ and the explicit form of these conditions read as
\begin{widetext}
\begin{eqnarray}
&&
 4 (a \gamma^2  {\cal L}-2 {\cal E} r)^2-\frac{r^2-2 r +a^2 }{\gamma ^4} \bigg[4 \gamma^8 {\cal L}^2-8 a^2 \epsilon^2 k \gamma^2  r+\epsilon^2 r \left( 2 a^2+a^2 r+r^3\right)
\nonumber \\ &&
\qquad -4 \gamma^4 r \left((a \epsilon k+{\cal E})^2 r-\epsilon {\cal L}r-r+ 2 {\cal E}^2- 2 a^2 \epsilon^2 k^2 \right)\bigg] =0\ , \label{con1} \\
&&
2 \epsilon ^2 \left(a^4 r+3a^4 -4 a^2 r-r^5+ r^4\right)-8 a^2 \gamma ^2 k \epsilon ^2 (r^2-4r+3 a^2)+8 \gamma ^4 \bigg[2 a {\cal E} k r^2 \epsilon-a^4 k^2 \epsilon^2 (r-3) -2 a^3 {\cal E} k r \epsilon
\nonumber \\
&&
\qquad +a^2 \left(2 k^2r^2 \epsilon^2 -4 k^2r  \epsilon^2 +{\cal L} r \epsilon+r- {\cal E}^2 r-3  {\cal E}^2  \right) -r^2 ({\cal L} \epsilon +1)\bigg]+48 a \gamma^6 {\cal E} {\cal L}+8 \gamma^8 {\cal L}^2 (r-3) = 0\ , \label{con2}\\
&&
6 a^4 \epsilon ^2 \left(2 \gamma^2 k-1\right) \left(2 \gamma^2 k r-8\gamma^2 k+r+4\right) +8 a^2 \bigg[3 \gamma^4 \left({\cal E}^2 r+4 {\cal E}^2-r\right)-r \epsilon ^2 \left(2 \gamma^2 k-1\right) \left(2 \gamma^2 k r-6 \gamma^2 k+3\right)
\nonumber \\&&
\qquad  -3 \gamma^4 {\cal L} r \epsilon \bigg] +48 a^3 \gamma^4 {\cal E} k r \epsilon-32 a \gamma^4 {\cal E} \left(k r^2 \epsilon +6 \gamma^2 {\cal L}\right)-24 \gamma^8 {\cal L}^2 (r-4)+16 \gamma^4 r^2 ({\cal L} \epsilon +1)-2 r^5  \epsilon ^2\leq0\ . \label{con3}
\end{eqnarray}
\end{widetext}
If one considers the nonrotating boosted black hole in the absence of the magnetic field the consitions (\ref{con1})-(\ref{con3}) gives the result for ISCO radius as $r_{ISCO}/M =6$, which does not depend on the boost velocity. The boost velocity affects only on angular momentum of the particles, while energy of the particle on the circular orbits remains the same as in the case of Schwarzschild black hole.

\begin{figure*}[t!]
\begin{center}
\includegraphics[width=0.32\linewidth]{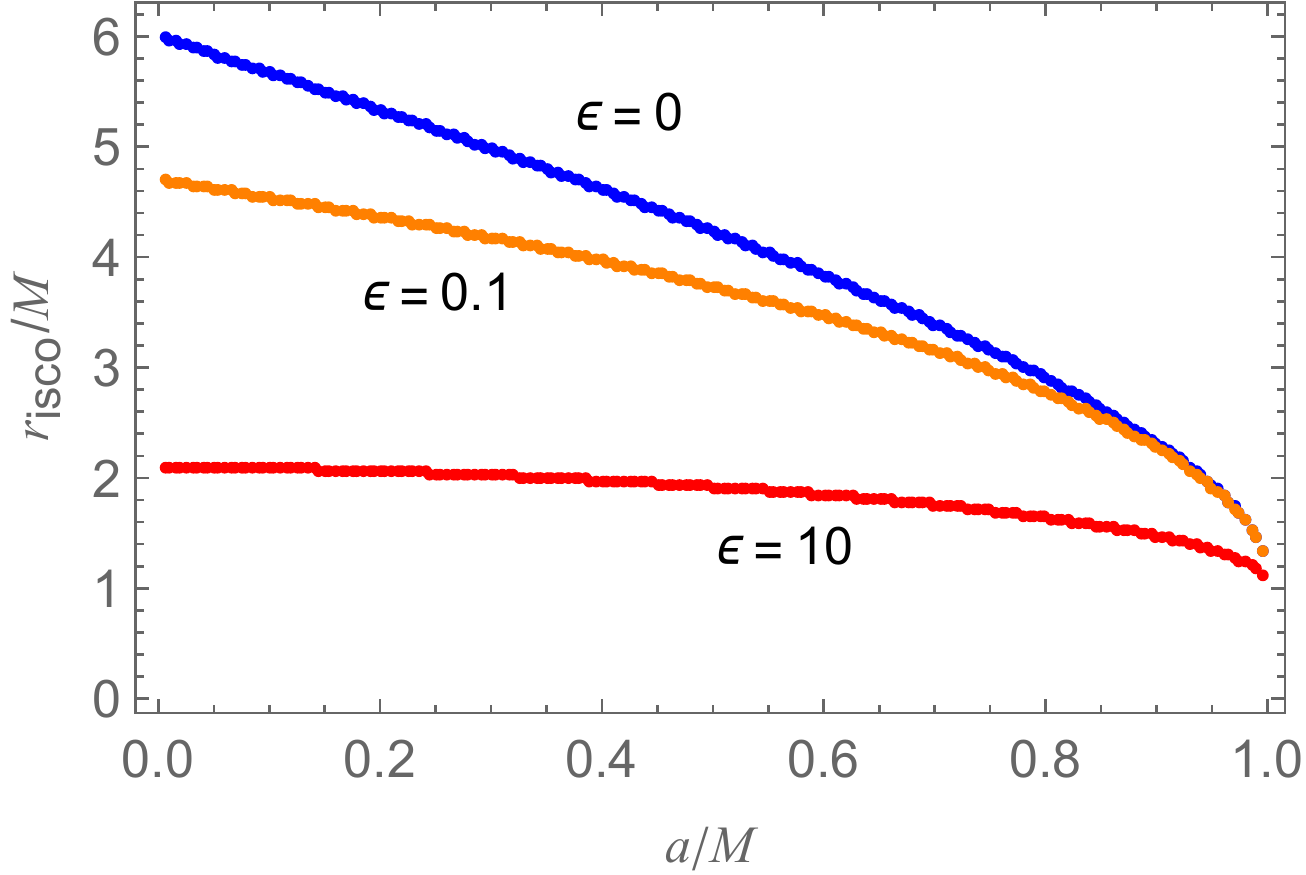}
\includegraphics[width=0.32\linewidth]{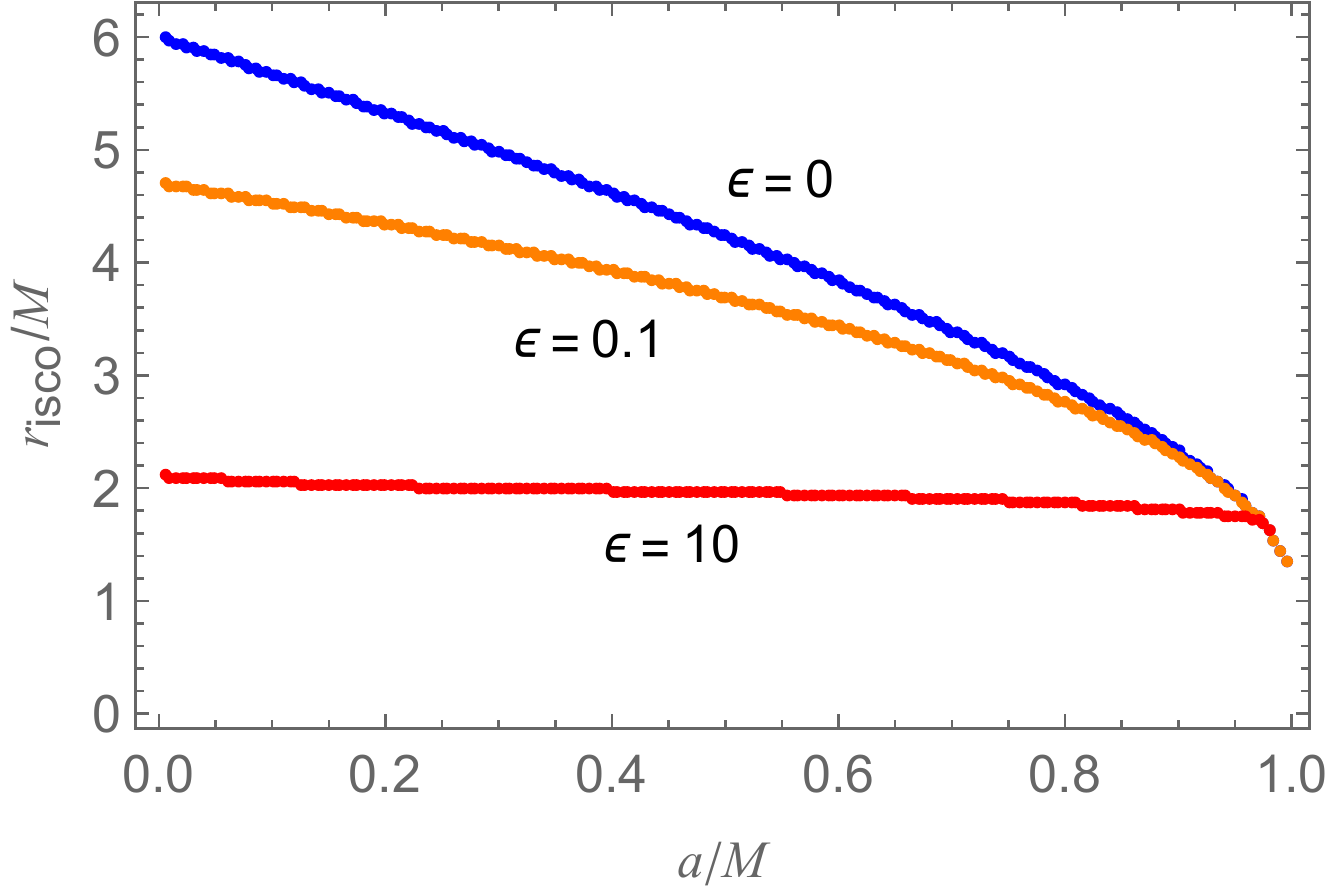}
\includegraphics[width=0.32\linewidth]{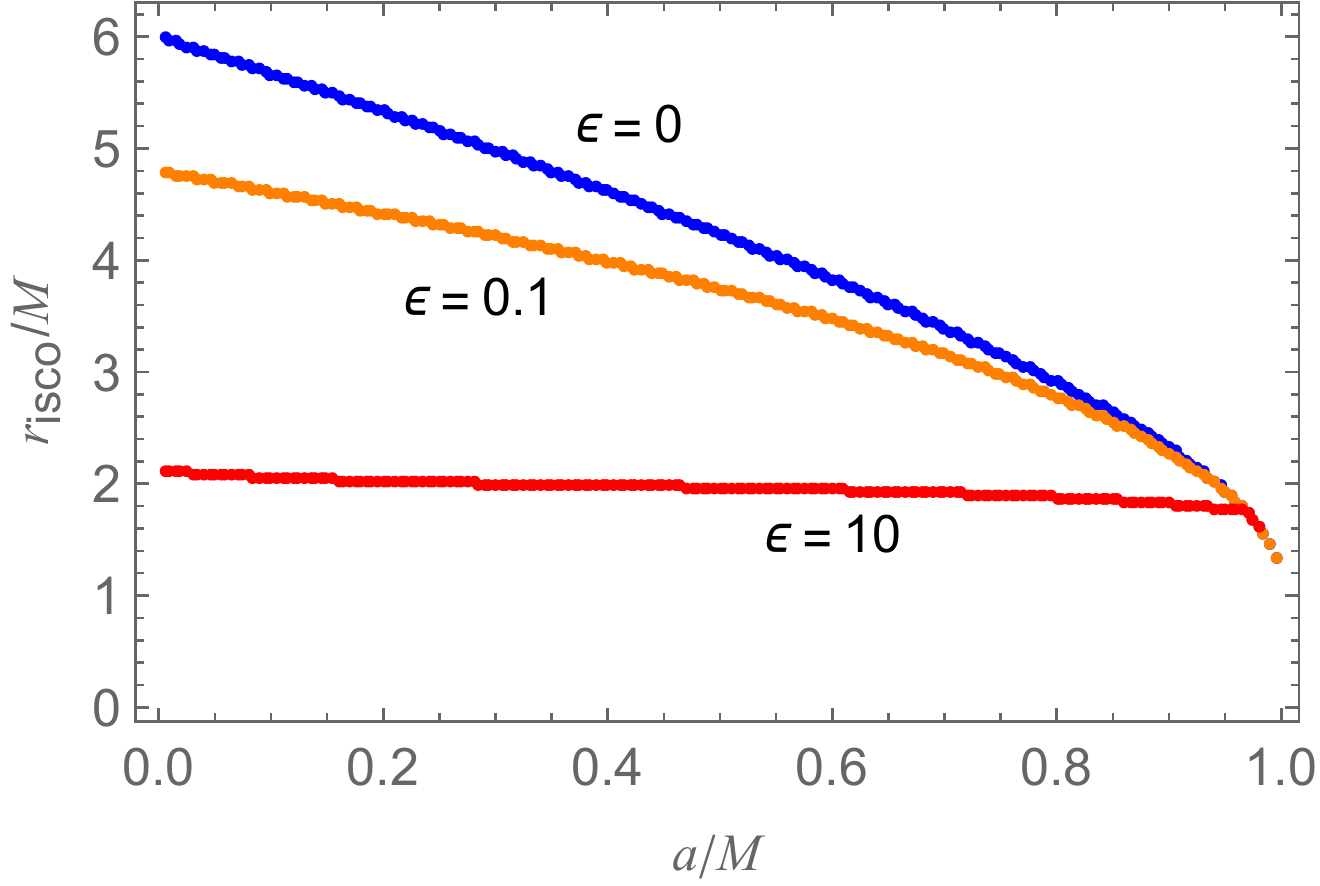}

\end{center}
\caption{ The dependence of the innermost stable circular orbits from the rotation parameter $a$ for the different values of magnetic parameter $\epsilon$ and boost velocity $v$: from the left to right $v=0; \ 0.5; \ 0.9$.  \label{riscop}}
\end{figure*}

\begin{figure}[t!]
\begin{center}
\includegraphics[width=0.9\linewidth]{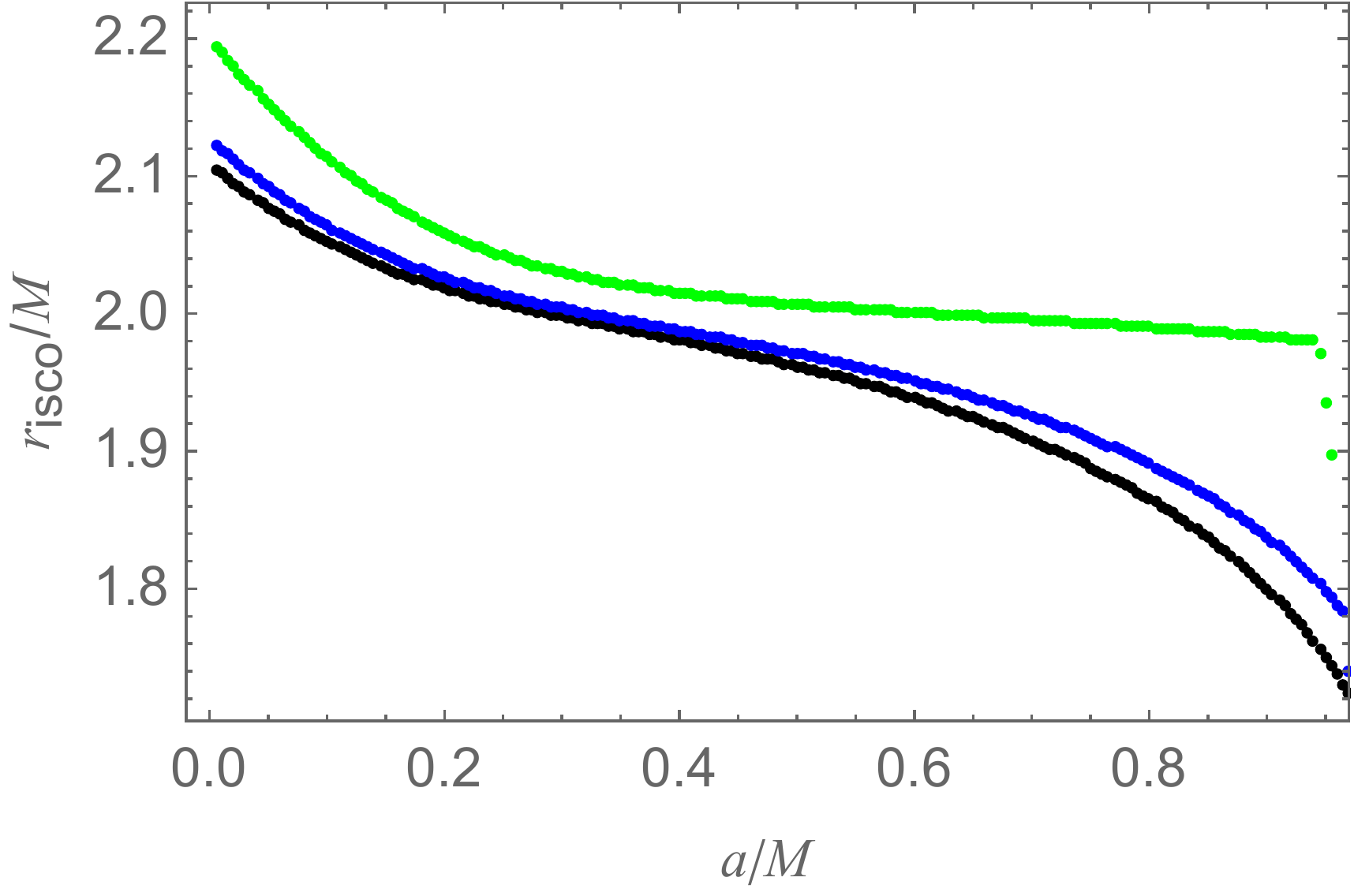}

\end{center}
\caption{The dependence of the innermost stable circular orbits from the rotation parameter $a$ for the different values of  boost velocity $v$: $v=0.1$ (black), $v=0.5$ (blue), $v=0.9$ (green),  for the fixed value of the magnetic parameter $\epsilon=10$.   \label{riscov}}
\end{figure}

The numerical solution of these set of equations has been shown  in Figs.~\ref{riscop} and \ref{riscov} and they depict variation of ISCO radius with rotation. The former refers to different values of $\epsilon$ and boost velocity while for the latter $\epsilon=10$ is fixed. ISCO radius decreases with increasing rotation. As $\epsilon$ increases the initial no rotation value $r/M=6$
decreases and then it goes down to $r/M=1$ for extremally rotating black hole $a/M=1$.


%

\section{Energetics of rotating boosted black hole  \label{sect4}}

In this section we would employ magnetic Penrose process (MPP) for extracting rotational energy of boosted Kerr black hole. In the original Penrose process it is envisaged that a particle
falling from infinity splits into two fragments in the ergoregion, one of which attains negative energy and falls into the black hole while the other would by conservation of energy come out with energy greater than that of the incident particle. It was soon shown \cite{Wald74,Wald74c,Bardeen72} that for a fragment to ride on negative energy orbit, the relative velocity between the fragments $>1/2 c$ which is astrophysically unsustainable. Then in Ref. \cite{wagh85}, it was proposed that a rotating black hole sits in a magnetic field produced due to the currents in accretion disk, and so a neutral particle splits into two charged fragments. Now the energy required to get onto negative energy orbit could come from electromagnetic interaction leaving velocity between fragments completely free. This was how PP was revived as MPP for astrophysical applications. We shall here consider the effect of boost velocity on the efficiency of this process.

MPP is the mechanism with the help of which the energy may be extracted from a rotating black hole in the presence of the external magnetic field  in an astrophysical setting. Astrophysical rotating black holes and AGNs always have accretion disk surrounding them and that would produce a magnetic field. This is how a proper environment is created for MPP to be fully operative.

\begin{figure}[t!]
\begin{center}
\includegraphics[width=0.9\linewidth]{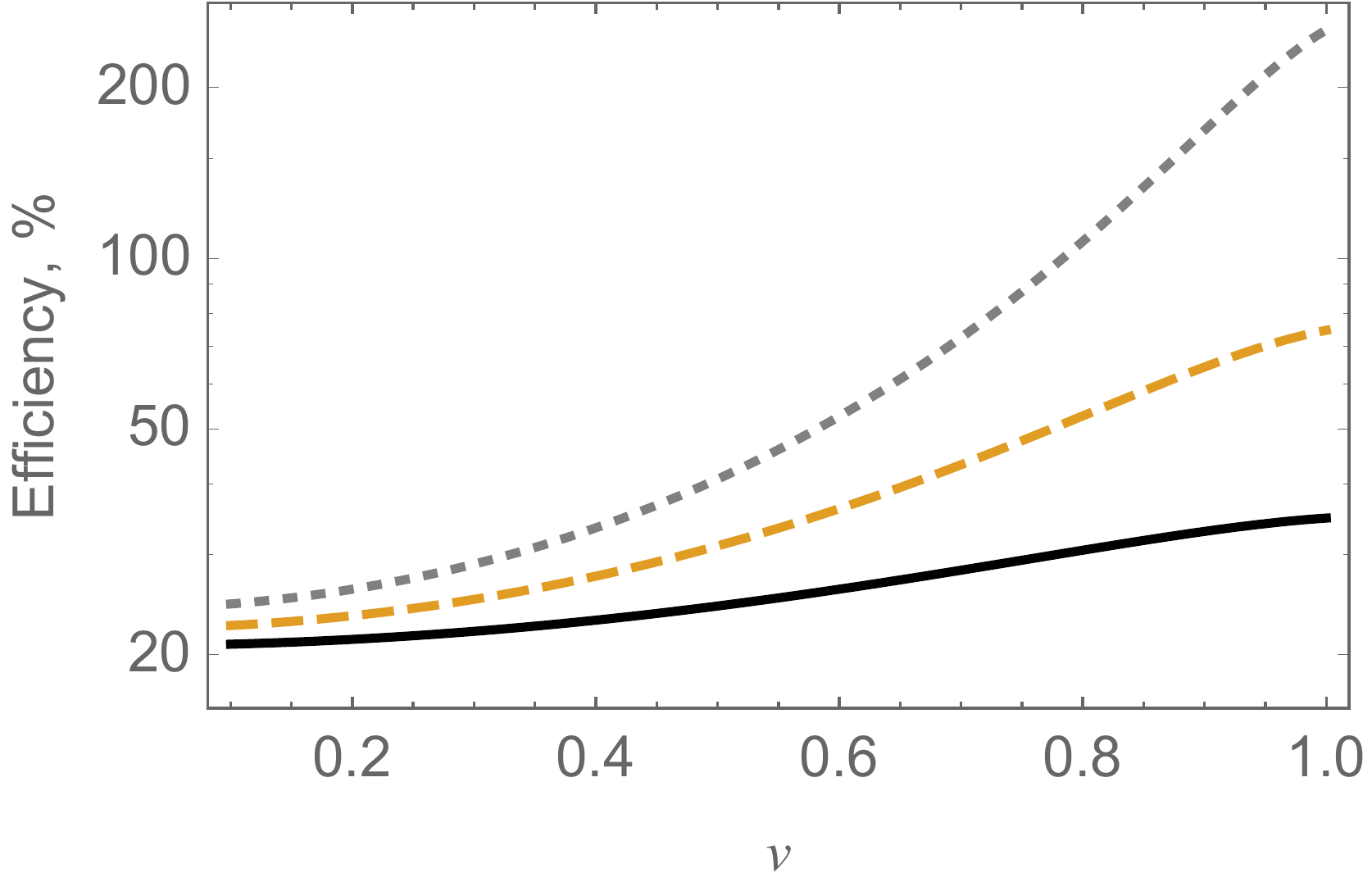}

\end{center}
\caption{The dependence of the energy extraction efficiency on magnetic parameter for the different values of the magnetic parameter: $\epsilon=0$ (solid line), $\epsilon=0.1$ (dashed line), $\epsilon=1$ (dotted line).  \label{efficiency}}
\end{figure}

Using the standard way definition of energy extraction efficiency due to Penrose process one may write 
\begin{eqnarray}
\eta=\frac{{\cal E}_2-{\cal E}_1}{{\cal E}_0}\ ,
\end{eqnarray}
where ${\cal E}_0,\ {\cal E}_1, \ {\cal E}_2$ are the energies of the initial, captured and released partiles, respectively. Using the energy and angular momentum conservation laws one can find the expression for the efficiency as
\begin{eqnarray}
\eta&=&\frac{\sqrt{2 \Lambda^2 M r -r^2 \Lambda^2 +\kappa^2}}{2 \kappa} - 1/2\, , \\
         \kappa&=&\Lambda r +
      a \epsilon \left(1 - 2 k \Lambda + k \Lambda \frac{r}{M}\right)\ ,
\end{eqnarray}
and for the extreme rotating boosted Kerr black hole as
\begin{eqnarray}
\eta= \frac{\sqrt{\Lambda^2 + (\Lambda + a \epsilon (1 - k \Lambda))^2}}{
 2 \sqrt{(\Lambda + a \epsilon (1 - k \Lambda))^2}} - \frac{1}{2}\ .
\end{eqnarray}
In the Fig.~\ref{efficiency} the dependence of energy extraction efficiency on magnetic parameter is shown for different values of the boost velocity.  From this dependence one can see that the magnetic field and boosted velocity coupling will increase the energy efficiency. However as noted earlier while defining the ergoregion, increase in boost velocity results in shrinking of the ergoregion. The effect of boost velocity is rather mixed in the sense that it tends to increase the efficiency of the process while at the same times the region where the process takes place gets shrunk. In the absence of both magnetic field and boost, the maximal value of the efficiency of the energy release is $20.7 \%$ of the Kerr blck hole.

\section{Conclusion \label{sect6}}

In this paper, we have studied the electromagnetic field structure and energetic process around boosted rotating black hole. The obtained results of the paper can be summarized as follows:

\begin{itemize}

\item We have found the exact analytic expressions for the potential and electromagnetic field components for a boosted rotating black hole immersed in an external asymptotically uniform magnetic field. The analysis shows that electromagnetic field structure around the black hole is sensitive to boost velocity.  Magnetic field is asymptotically uniform in which black hole is sitting while electric field is quadrupolar produced by twisting of magnetic field lines, and hence decays very fast. In comparison, electric field is relatively more sensitive to boost velocity than magnetic field. The effect of velocity is to reorient field lines in relation to the boost direction. In the asymptotic infinity the components of both electric and magnetic field tend to their Newtonian expressions.

\item Charged particle motion around the boosted rotating black hole has been studied, and we have employed the Hamilton-Jacobi equation and considered motion confined to the equatorial plane. The effective potential profile will be affected by boost velocity mainly in the near region while the effect of the magnetic field picks up in the far region. 
    
\item It turns out that boost velocity has no effect on the radius  of the innermost stable circular orbit which however decreases with increasing $a/M$ as well as $\epsilon$. 

\item The magnetic Penrose process around boosted rotating black hole has been considered. The exact analytic expression for the efficiency of magnetic Penrose process in the presence of boost velocity  has been obtained. It was shown that boost velocity increases the efficiency of the magnetic Penrose process. On the other hand it tends to shrink the width of ergoregion which is actually the playground for the process. Its overall effect is rather mixed blessing -- the efficiency increases but the region of action gets diminished.

\end{itemize}

\section*{Acknowledgment}
B.A. and A.A. acknowledge the Inter-University Centre for Astronomy and Astrophysics, Pune, India  and the Goethe University, Frankfurt am Main, Germany for
warm hospitality, and ND thanks AEI, Potsdam-Golm for the same. This research is supported in part by
Projects No.~VA-FA-F-2-008, No.~YFA-Ftech-2018-8 of
the Uzbekistan Ministry of Innovation Development, and by the Abdus Salam International Centre
for Theoretical Physics through Grant No.~OEA-NT-01 and by the Volkswagen Stiftung, Grant No.~86 866. This research is partially
supported by Erasmus+ exchange grant between SU and NUUz.  A.A. acknowledges the TWAS associateship programm for the support.

\bibliographystyle{apsrev4-1}  
\bibliography{gravreferences}

\end{document}